\documentclass[11pt]{article}

\usepackage{times}
\usepackage{graphicx}
\usepackage{amssymb,amsmath}
\usepackage{multirow}
\usepackage{caption}
\usepackage{threeparttable}
\usepackage [latin1]{inputenc}
\usepackage{natbib}

\renewcommand\refname{References}

\newcounter{lastnote}

\title{A New Position Calibration Method for MUSER Images}

\author
{Zhichao~Zhou$^{1,2,3}$, Yihua~Yan$^{1,2,3\ast}$, Linjie~Chen$^{1,2}$, Wei~Wang$^{1,2}$, Suli~Ma$^{1,2}$\\
\footnotesize{$^{1}$State Key Laboratory of Space Weather, National Space Science Center, Chinese Academy of Sciences,  Beijing 100190, China}\\
\footnotesize{$^{2}$National Astronomical Observatories, Chinese Academy of Sciences, Beijing,100101, China}\\
\footnotesize{$^{3}$School of Astronomy and Space Sciences, University of CAS, Beijing 100049, China}\\
\footnotesize{$^\ast$Corresponding author; E-mail:  yanyihua@nssc.ac.cn, yyh@nao.cas.cn.}
}

\date{Accepted by {\it Research in Astronomy and  Astrophysics} on Aug. 12, 2022}

\begin{document}


\baselineskip18pt


\maketitle


\abstract{
The Mingantu Spectral Radioheliograph (MUSER),  a new generation of solar dedicated radio imaging-spectroscopic telescope, has realized high-time, high-angular, and high-frequency resolution imaging of the sun over an ultra-broadband frequency range. Each pair of MUSER antennas measures the complex visibility in the aperture plane for each integration time and frequency channel. The corresponding radio image for each integration time and frequency channel is then obtained by inverse Fourier transformation of the visibility data. However, the phase of the complex visibility is in general severely corrupted by instrumental and propagation effects. Therefore, robust calibration procedures are vital in order to obtain high-fidelity radio images. While there are many calibration techniques available -- e.g., using redundant baselines, observing standard cosmic sources, or fitting the solar disk -- to correct the visibility data for the above-mentioned phase errors, MUSER is configured with non-redundant baselines and the solar disk structure cannot always be exploited. Therefore it is desirable to develop alternative calibration methods in addition to these available techniques whenever appropriate for MUSER to obtain reliable radio images. In the case that a point-like calibration source containing an unknown position error, we have for the first time derived a mathematical model to describe the problem and proposed an optimization method to calibrate this unknown error by studying the offset of the positions of radio images over a certain period of the time interval. Simulation experiments and actual observational data analyses indicate that this method is valid and feasible. For MUSER's practical data the calibrated position errors are within the spatial angular resolution of the instrument. This calibration method can also be used in other situations for radio aperture synthesis observations.\\[2ex]

}


\section{Introduction}           
\label{sect:intro}
 For an interferometer, radio signals from a cosmic source are received by the antenna of an observation station, and then transmitted indoors to complete the processes of mixing, amplification and filtering. Then the signal is digitalized by the digital receiver. Afterwards, the complex signals of the two antennas are correlated to produce observational visibilities \citep{Thompson+2017}, equivalent to a Fourier component of the radio brightness distribution. The term "calibration" refers to the estimation and correction for the instrumental gain and errors in the visibilities \citep{Grobler+2014}. The purpose of calibration is to solve the unknown gain error and phase error of the equipment, as well as the unknown propagating interference \citep{Wijnholds+2010}. The available calibration methods of radio telescopes mainly include three basic categories: direct calibration, calibration referenced to calibrator sources in the sky, and self-calibration \citep{Bastian1989,Fomalont+Perley1999,Thompson+2017}. Direct calibration measures the amplitude gain and delay phase in the system link by constructing a link loop. The principle of observing the calibration source is to use the radio telescope to observe the target source and a known calibration source, and then remove the influence of the instrument and the propagation path based on the two observation data. The self-calibration method uses the characteristics of the target source and the characteristics of the antenna array, e.g., the closure relationships, to build a model to achieve the desired calibration result. The calibration and imaging methods can be treated within a common mathematical framework and the calibration is reduced to the mutual fitting of the observed values of sky model and instrument model, such as the model provided by the radio interferometry equation \citep{Hamaker+1996, Smirnov+2011, Rau+2009}.  Unresolved point sources are normally employed as calibrators because their phase closure should be zero and their amplitude closure unity. Thus, they are useful in checking the accuracy of calibration and examining instrumental effects \citep{Thompson+2017}. Future radio telescopes will have a large number of antennas and a large field of view. In this case, further considerations should be taken into account for calibrations, e.g., to deal with the parameters with a strong directional dependence, etc. \citep{Wijnholds+2010}.

 For the calibration of a radioheliograph, a radio telescopes designed to observe the sun, a redundant baseline design was adopted by the Nobeyama Radioheliograph \citep{Nakajima+1994}, the {Nancay} Radioheliograph \citep{NRH1993} and the Siberian Radioheliograph\citep{Altyntsev+2020}, where the number of redundant correlations is greater than the number of antennas.  The least square method is used to solve for antenna gains and to correct for phase errors of each antenna, based on the principle that the phases recorded by equal baselines should be the same \citep{Nakajima+1994, Altyntsev+2020}. When the upgraded Very Large Array (VLA, \citealt{Thompson+1980}) observes the sun, the complex gain and delay phase of the telescope system are calibrated by observing a standard cosmic source \citep{Chen+2012}. Since small antennas may have insufficient sensitivity to observe cosmic calibrator sources, the Expanded Owens Valley Solar Array (EOVSA, \citealt{Nita+2016}) introduced a 27-meter antenna equipped with He-cooled receivers, to calibrate the small antennas using standard calibrator source.

 The Mingantu Spectral Radioheliograph (MUSER), which is stationed in Inner Mongolia, China, is a new generation of radioheliograph capable of observing the Sun with high time, angular, and frequency resolution \citep{Yan+2009,Yan+2021}. Observations by MUSER during 2014-2019 have been presented by  \citet{Zhang+2021}. The outcomes of calibration and data processing for MUSER in decimetric wavelengths have been reported, including delay measurements, polarisation calibration, and some additional results of calibration and data processing in \citet{Wang+2013}.  The approach to calibrating MUSER is to observe the strong radio sources in the sky as the point-source calibrator. Since MUSER antennas are insensitive to cosmic calibrator sources, radio beacons on satellites are the strongest sources in the sky for MUSER. In the MUSER frequency band, some geosynchronous orbit satellites and GPS satellites are available. Therefore, the satellites can be observed as calibrator sources at several discrete frequencies. Some strong radio sources or intensive radio bursts on the sun can also be used as calibrator sources across the full frequency band \citep{Wang+2013, Wang+Yan2019,Wang+2019}. 

However, when a satellite is used as a calibrator, its nominal position may not be as accurate as celestial source's position in real time except for GPS or other navigation and positioning satellites, though the accuracy of this nominal position may still meet the needs of the satellite's original purpose for  applications.  This satellite nominal position contains an error which will cause a solar radio image to deviate from the center of the field of view. Fortunately, the solar disk in the solar radio image can be employed to determine the disk center by fitting the solar disk model so as to obtain the offset of the solar image \citep{Mei+2017,Chen+2017,Wang+2019}. Nevertheless, this approach may not work in general: the solar disk structure in a solar image is not always obvious due to sparse sampling of the synthesis array, or observing at low radio frequencies. Therefore, we have for the first time derived a mathematical model  to describe the problem with the calibrator position deviation, and proposed a new method to determine the calibrator's  position error by minimizing to zero the RMS error of the deviated positions away from the center of radio images over a certain period of time interval. 

  As described in \citet{Thompson+2017}, the closure phase for the point-source is always zero, even if it is not at the phase-tracking center or if the station coordinates have errors. Therefore the position of the point-source cannot be deduced from closure phase measurements alone. While a point-source is an ideal calibrator in radio interferometry and/or radio aperture synthesis, it turns out that its position cannot be determined by the aperture synthesis theory exclusively, but must be prescribed in advance by other means.  

In the next Section \ref{section2}, we briefly introduce the characteristics of MUSER.  This new mathematical model, as well as the resulting new position calibration method, are presented in Section~\ref{section3}. In Section \ref{section4} the simulation results are shown to validate the method. The calibration results of real MUSER observations are also demonstrated. Finally we discuss the merits of the new method and provide our conclusions in Section \ref{section5}.


\section{Brief Description of MUSER Imaging}
\label{section2}
\begin{figure}
 \centering
   \includegraphics[width=0.85\textwidth]{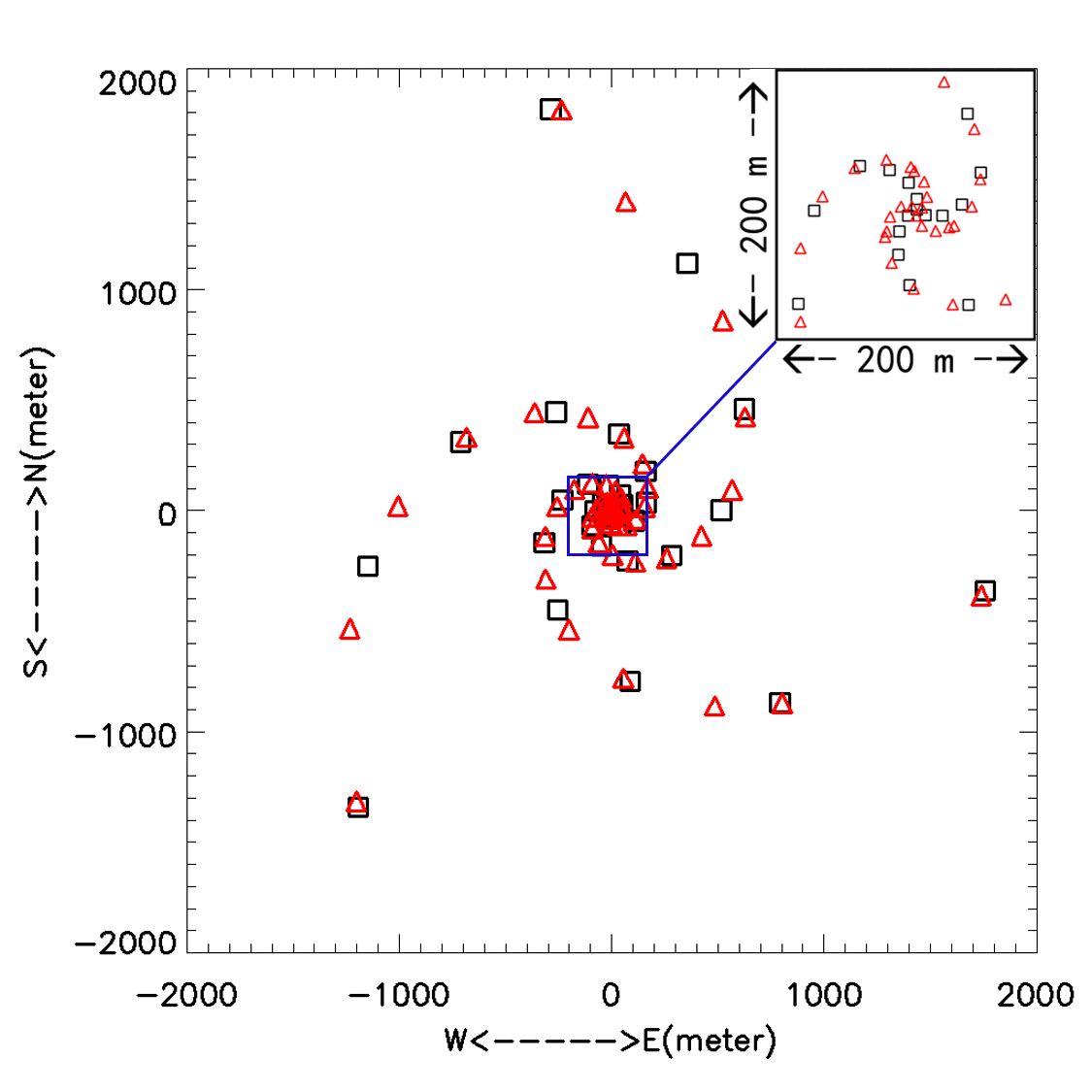}
    \caption{Antenna array configuration for MUSER with maximum baseline of $\sim$3 km. The inserted panel shows in detail the antenna locations in the central area within 200m range. Note that the black squares represent MUSER-I antennas; the red triangles denote MUSER-II antennas. The central antenna location is E $115^{\circ}15^{\prime}1.8^{\prime\prime}$ longitude and N $42^{\circ}12^{\prime}42.6^{\prime\prime}$ latitude, with the altitude of 1365m.}
    \label{antennas position}
\end{figure}

The main characteristics and performance of MUSER  are listed in Table~\ref{characterestics}\citep{Yan+2009,Yan+2021}. Presently MUSER consists of two arrays named as MUSER-I and MUSER-II.  MUSER-I contains 40 antennas of 4.5 m diameter operating in the frequency range from 0.4 GHz to 2 GHz, and MUSER-II contains 60 antennas of 2m diameter operating in the frequency range from 2 GHz to 15 GHz. These 100 antennas are arranged on three logarithmic spiral arms shown in Fig.~\ref{antennas position}, and the longest baseline is about 3km \citep{Yan+2009}. The inserted panel in Fig.~\ref{antennas position} shows the dense antenna distribution in the central area within 200m range.  

\begin{table}
 \caption{MUSER characterestics and performance.}
 \label{characterestics}
 \begin{tabular*}{\columnwidth}{l@{\hspace*{22pt}}l@{\hspace*{22pt}}l}
  \hline
  MUSER Array & MUSER-I & MUSER-II\\
  \hline
  Frequency range: & 400 MHz - 2 GHz & 2  - 15 GHz\\[2pt] 
  Array antennas & 40 $\times \phi$4.5 m & 60 $\times\phi$2 m\\[2pt]
  Single dish beam: & 9.5$^\circ$ - 1.9$^\circ$ &4.3$^\circ$-$0.6^\circ$\\[2pt]
  Frequency resolution: & 64 channels & 520 channels\\[2pt]
  Angular resolution: & 51.6$^{\prime \prime}$- 10.3$^{\prime \prime}$ & 10.3$^{\prime \prime}$- 1.3$^{\prime \prime}$\\[2pt]
  Time resolution: & 25 ms & 206.25 ms\\[2pt]
  Dynamic range: & 25 db (snapshot)  & 25 db (snapshot)\\[2pt]
  Polarizations: & Dual circular L, R & Dual circular L, R\\[2pt]
  Maximum baseline: & $\sim$3 km & $\sim$3 km\\[2pt]
  \hline
 \end{tabular*}
\end{table}

As an aperture synthesis radio telescope, MUSER images the sun by following the standard aperture synthesis imaging process in radio astronomy. However, there are also some specific characteristics in MUSER imaging. A high-performance imaging pipeline and some algorithms have been developed for MUSER to produce solar radio images \citep{Mei+2017,Chen+2017,Wang+2019,Chen+2019}. E.g., We have done some research work in deconvolution of extended source, like the sun.  Deconvolution using the generated countermeasure network is proposed \citep{Xu+2020}, and the simulation results show that it is better than theHogbom CLEAN algorithm \citep{Thompson+2017}. Another, the Cornwell Multi-scale CLEAN \citep{Cornwell2008} has been employed as the deconvolution method in MUSER solar radio imaging \citep{Zhao+2017}. The image structure information were obtained by combining the weighting functions of natural weighting and uniform weighting \citep{Wang+Yan2019}. The quasi-periodic pulsations before and during a solar flare were analyzed with restored radio images observed by MUSER \citep{Chen+2019}, etc. 

 Since the two MUSER arrays have no redundant baselines, the calibration procedure utilised in other radioheliographs to account for redundant baselines cannot be employed for MUSER. It is also difficult for MUSER to observe weak radio source signals in the sky as MUSER arrays are composed of small antennas. The known strong radio sources as calibrators are needed for the phase calibration. Therefore geosynchronous satellites such as meteorological satellites with strong signals have been used as the calibrator sources for MUSER with spherical wavefront effects due to the small distance of the satellites from the Earth have also been taken into account \citep{Wang+2019}. 

 As mentioned above, though, while the nominal position of the satellite may satisfy its purpose, it may be offset from the actual position of the satellite at the time of measurement that cause issues for source positions. Tracking a satellite in 9 hours reveals that the satellite position errors are  $\sim1.1^{\circ}$ in Declination and $\sim1^{\prime}$(arcmin) in Hour angle \citep{Wang+2019}. 
This satellite position error will cause solar radio images obtained by MUSER to deviate from the center of the field of view. Therefore other methods such as fitting the solar disk model have been applied to correct the offset of the solar images \citep{Mei+2017,Chen+2017,Wang+2019}. Within the framework of radio interferometry and aperture synthesis theory, we present a new method for determining the position error of the satellite (or calibrator), i.e., determining this unknown deviation from the phase-tracking center.

\section{Method of Determining Calibrator's Position Deviation}
\label{section3}

 In the following subsections, we derive the mathematical basis for calibration with a point-source calibrator offset from the phase tracking center and describe the numerical procedure used to solve the problem.

\subsection{Calibration with a source offset from the phase tracking center}
\label{subsect3.1}

Under the approximation that the synthesized field of view is small, the radio distribution of brightness on the sky can be obtained from the following inverse Fourier Transform:
\begin{equation}
  I^D(l,m)=\iint S(u,v) V(u,v)e^{j2\pi(ul+vm)} dudv,
	\label{eq:dirty}
\end{equation}
where $V(u,v)$ is called the visibility function and $S(u,v)$ is the sampling function produced by all baselines over ($u$,$v$) plane. The visibilities $V(u,v)$ are obtained through the correlation of radio signals of any two antennas in the interferometric array, correspond to the Fourier components of the radio intensity distribution in the observed small sky area. $I^D(l,m)$ is the so-called dirty image. 

The measured visibilities must be calibrated to remove the influence of the instrumental gain and other effects. This can be achieved by observing a calibrator source. When pointing at the calibrator position, or the phase tracking center, it is clear that all the visibility amplitudes are unity and all the visibility phases are zero if the calibrator is a point-source of unit flux density. Hence, the complex gain of each antenna can be determined \citep{Bastian1989,Fomalont+Perley1999}. Normally an interferometric array is designed to be stable during observations and the phase errors corresponding to each interferometric element are assumed stochastic stationary when pointing at different directions. 

When pointing at the calibrator source, the phase tracking center is set to the calibrator position and it is at the origin of the image plane $(l,m)$. However, if there exists an offset term ($l_d,m_d$) from the origin (i.e., the nominal position of the calibrator, or the phase tracking center),  the actual observed visibilities  for the calibrator become:
\begin{equation}
V_{cal}(u,v)e^{j2\pi(ul_d+vm_d)}=\iint I_{cal}(l+l_d,m+m_d)e^{-j2\pi(ul+vm)}dldm,
 \label{eq:basecal}
\end{equation}
and the observed response of the interferometric element with antenna pair ($p$-$q$) for the calibrator source is:
\begin{equation}
    {r_{cal}^{pq}}=|V_{cal}^{pq}|g^{pq}\exp[{-j(\varphi_{Vcal}^{pq}+\varphi_{err}^{pq})+j2\pi(u_{cal}^{pq}l_d+v_{cal}^{pq}m_d)}],
	\label{eq:calerr}
\end{equation}
where $u_{cal}^{pq},v_{cal}^{pq}$ are the corresponding $u,v$ coordinates when pointing at the calibrator source for the antenna pair ($p$-$q$),   
and $\varphi_{err}^{pq}$ is the  phase error of the corresponding interferometric element due to instrumental effects. This deviation will be transferred to the target radio images through the phase calibration process with a point calibrator source. The calibrated response of the interferometric element with antenna pair ($p$-$q$) for the solar image becomes \citep{Wang+2019}:
\begin{equation}
    {r_{sun}^{pq}}=|V_{sun}^{pq}|g^{pq}\exp[{-j\varphi_{Vsun}^{pq}+j2\pi(u_{cal}^{pq}l_d+v_{cal}^{pq}m_d)}], 
	\label{eq:sunerr}
\end{equation}
where  $|V_{sun}^{pq}|$ indicates the visibility amplitude of the sun and $\varphi_{Vsun}^{pq}$ is the corresponding visibility phase of the sun. 
The solar visibilities thus calibrated with a calibrator offset from the phase tracking center can then be expressed as follows.
\begin{equation}
V(u,v)=V_{sun}(u,v)e^{j2\pi(u_{cal}l_d+v_{cal}m_d)}=V_{sun}(u,v)e^{j2\pi[u(\frac{u_{cal}}{u}l_d)+v(\frac{v_{cal}}{v}m_d)]}.
 	\label{eq:solvis}
\end{equation}

When pointing at the calibrator,  $u_{cal},v_{cal}$ are different for different baselines. We can decompose the baseline ratios when pointing at the calibrator source and the target objective, the sun, as an invariant term for all baselines and a variable function as follows, 
\begin{equation}
\frac{u_{cal}}{u}=\xi_0+\frac{\xi(u,v)}{u}~,~~~\frac{v_{cal}}{v}=\eta_0+\frac{\eta(u,v)}{v},
 	\label{eq:deviaterm}
\end{equation}
in which $\xi_0, \eta_0$ are constants and $\xi(u,v),  \eta(u,v)$ are functions of $u, v$ because a given baseline in general traces an ellipse in the (u,v) plane, {with hour angle as the variable} \citep{Fomalont+Perley1999,Thompson+2017}. Therefore, {$\xi_0, \eta_0$} are also changing with time although they are constants for a certain instant.  

By substituting equation~(\ref{eq:solvis}) into equation~(\ref{eq:dirty}) with the decomposed expression~(\ref{eq:deviaterm}) and carrying out the inverse Fourier Transform, we find that the final dirty image of the observation {with {$l_d,m_d$} as parameters} is the deviated solar image convolved with a {blurring} function. 
\begin{equation*}
I^D(l,m;l_d,m_d)= \iint S(u,v) V_{sun}(u,v)e^{j2\pi[u(\xi_0l_d)+v(\eta_0m_d)]}\times e^{j2\pi[\xi(u,v)l_d+\eta(u,v)m_d]}e^{j2\pi(ul+vm)} dudv
\end{equation*}
\begin{equation}
\null~ \hspace{-3.6cm}= I_{sun}^d(l+\xi_0l_d,m+\eta_0m_d)*H(l,m;l_d,m_d),
 	\label{eq:sunshift}
\end{equation}
in which the solar image $I_{sun}^d$ deviated from the phase tracking center is 
\begin{equation}
I_{sun}^d(l+\xi_0l_d,m+\eta_0m_d)
=\iint S(u,v) V_{sun}(u,v)e^{j2\pi[u(\xi_0l_d)+v(\eta_0m_d)]} e^{j2\pi(ul+vm)} dudv, 
 	\label{eq:basesun}
\end{equation}
and {$H(l,m;l_d,m_d)$} is defined formally by the following expression,
\begin{equation}
H(l,m;l_d,m_d)=\iint  e^{j2\pi[\xi(u,v)l_d+\eta(u,v)m_d]}e^{j2\pi(ul+vm)} dudv.
 	\label{eq:blur}
\end{equation}
Though it is difficult to obtain the analytic expression of {$H(l,m;l_d,m_d)$},  we may nevertheless estimate its influence on the final dirty image of the sun as a {blurring} effect.  From the Fourier Transform function of the form $e^{-j\phi}$ in the integrand in equation~(\ref{eq:blur}) we can see that its modulus is 1. Therefore {{$H(l,m;l_d,m_d)$}} may modulate the solar dirty image distribution without changing the maximum intensity of the original dirty image and total energy of the original signal. Furthermore, in the case that $l_d=0$ and $m_d=0$, the Fourier Transform function in equation~(\ref{eq:blur})  becomes a constant 1.  It turns out that the inverse Fourier Transform {{$H(l,m;l_d,m_d)$}} becomes a $\delta$-function under the condition of 
$l_d=0$ and $m_d=0$. Equation~(\ref{eq:sunshift}) is derived for the first time to address the problem of calibrator position deviation from the phase tracking center, and it provides a solid mathematical foundation for our new position calibration procedure.

 In practice, the dirty image obtained after calibration with a satellite signal is the target dirty image shifted with the deviation ($\xi_0l_d$,$\eta_0m_d$) and modulated by the above expressed blurring function {$H(l,m;l_d,m_d)$}. As mentioned before, this shifting amount is also changing with time. The position error ({$l_d, m_d$}) of the satellite at the time of calibration is largely unknown, but it has fixed values. 
Therefore the phases  {$2\pi$}{$(u_{cal}l_d$}+{$v_{cal}m_d)$} introduced to corresponding baselines by the satellite position deviation for phase calibration are also unknown. Furthermore they are  inseparable from the corresponding visibility phase terms. Normally, it is difficult to directly obtain the exact position error of the satellite, or the calibrator. 

 Our strategy to eliminate the influence of the unknown position error ({$l_d, m_d$}) of the calibrator is based on introducing a deviation compensation ({$l_a, m_a$}) into the observed target visibility phases with equation ~({\ref{eq:sunshift}}). We employ the minus to express the expectation that the added phases {$2\pi$}{$[u({u_{cal}}/{u})l_a$}+{$v({v_{cal}}/{v})m_a]$} will act with opposite signs to the unknown calibrator's deviation with {$\Delta l=$}{$l_d-$}{$l_a$}, and {$\Delta m=$}{$m_d-$}{$m_a$}. Then the dirty image of the observation is modified as follows. 
 \begin{equation}
I^D(l,m;\Delta l,\Delta m)  =I_{sun}^d(l+\xi_0\Delta l,m+\eta_0\Delta m)*H(l,m;\Delta l,\Delta m),
 \label{eq:modelshift}
\end{equation}
 The above expression~({\ref{eq:modelshift}}) is the mathematical basis for our new position calibration procedure. If we could eliminate the influence of ({$l_d, m_d$}) by adjusting  ({$l_a$,$m_a$}) in the modified image {$I^D$}{$(l,m;$}{$\Delta l,\Delta m)$} with {$\Delta l$}, and {$\Delta m$} approaching  zero, we can obtain the desired final result. Of course if one has {$a~priori$} knowledge of ({$l_d, m_d$}) one can directly remove the influence of ({$l_d, m_d$}) by compensating {$l_a=$}{$l_d$} and {$m_a=$}{$m_d$}.  In general, we do not  know the exact deviation of the satellite calibrator at the time of the calibration. Therefore, we need a criterion to judge whether {$\Delta l$}=0 and {$\Delta m$}=0, which will be presented in the next subsection. 

\subsection{Procedure of new position calibration technique}
\label{subsect3.2}

\begin{figure}
  \centering
			\includegraphics[width=0.85\textwidth]{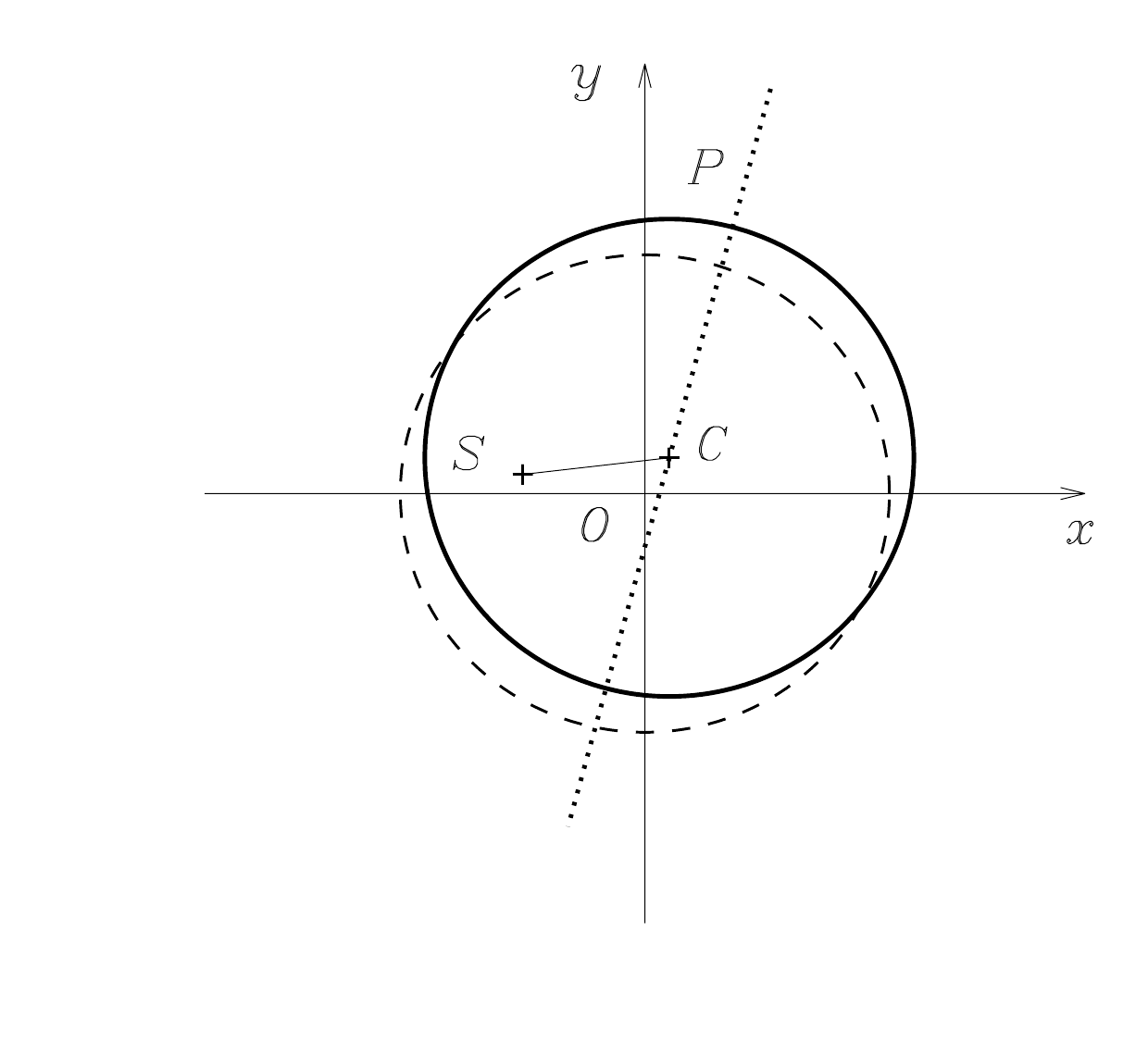}
		 \caption{A schematic plot displaying the image position deviation. The image coordinates are expressed in the plane of the sky, and $O$ is the phase tracking center with dashed circle indicating the desired solar location.  $C$ is the center of the solar disk indicated by the solid circle and $P$ is the projection angle of the sun on the sky plane between the solar rotation axis (indicated by the dotted line) and the north direction. $S$ is the position of a reference source on the solar disk.}
    \label{deviation}
\end{figure}

 As mentioned above, the solar visibility data calibrated by a calibrator with the unknown offset ({$l_d, m_d$}) relative to the phase tracking center, together with an arbitrarily added compensation ({$l_a$,$m_a$}), will generally cause the recovered solar radio image to deviate from the center of the field of view by the unknown offset ({$\xi_0$}{$\Delta l$}, {$\eta_0$}{$\Delta m$}). Fig.~{\ref{deviation}} schematically shows the location of the dirty image in which the solar disk center $C$ is offset from the phase tracking center $O$ with the corresponding unknown offset in the projected two-dimensional sky plane.
In this paper we do not consider the situation where the the solar disk can be fitted for the position calibration, which can nevertheless be applied to verify the calibrated results wherever appropriate.  $S$ denotes a reference source position on the solar spherical surface but projected on the two-dimensional solar disk. 
 
Apparently, if we adjust the added compensation ({$l_a$,$m_a$}) for the current observation, the solar image including the reference source $S$ and the center $C$ will also change accordingly. It should be noted that if the calibrator deviates from the phase tracking centre in an unexpected way, all locations on the target image will also deviate from the phase tracking centre with unknown offsets. The problem is determining the appropriate compensation in this situation.

If one has {\it a priori} knowledge of the location of the reference radio source on the solar disk, one can determine the actual offset of the satellite or calibrator by adjusting the compensation over just one solar image. However, this is not the case for the general calibration problem as considered here, i.e., we make no assumptions regarding the target solar images and the unknown calibrator offset relative to the phase tracking center. We instead consider the target images during a certain period of time. During this period, we may expect three kind of motions of a radio source in the target images. 

Firstly, the source moves with time in the same way as the center does in the target solar image due to the influence of unknown deviation errors ($\Delta l$,$\Delta m$) as described by equation~({\ref{eq:modelshift}}). Secondly, the source should rotate as the sun rotates with its axis in addition to the first kind of motion. Finally, the source may have its own relative motion on the solar surface in addition to the solar rotation and the first kind of motion. In all three cases, the source location will be modulated by the blurring function {$H(l,m;l_d,m_d)$}. 

If the reference source is a stable structure on the sun during a period of observation, the third kind of position variation for this source should be absent. Then the position change of the stable reference source with respect to the solar center is solely related to the rotation of the sun in addition to the influence of the unknown deviation error. Our goal is to eliminate the influence of the first kind of motion from the second kind of motion for a stable source. 

As shown in Figure~{\ref{deviation}}, while the unknown deviation error ($\Delta l$,$\Delta m$) influences both $S$  and $C$, the distance variation between $S$  and $C$ should be only due to  the rotation of the sun if the reference source is a stable structure on the sun during the observation time. This serves as our criterion for the new calibration procedure.

 As mentioned above,  the solar disk center $C$, which is in general unknown,  varies with respect to the phase center $O$ for images observed at different times due to the time-dependence of ({$\xi_0,\eta_0$}) in equation~{(\ref{eq:deviaterm})}  if the position error ($\Delta l$,$\Delta m$) has not been eliminated. Consequently, a source position $S$  on the solar disk also varies with respect to the phase center $O$ due to the same reason of the existence of the error term ($\Delta l$,$\Delta m$). We need to verify whether the difference between ${\bf R}_{sc}$ and  ${\bf R}_s$ decreases by adjusting the compensation deviation ($l_a, m_a$) from the calibrator phase tracking center. If  the compensation deviation matches the unknown deviation ($l_d, m_d$) of the calibrator from the phase tracking center, the difference should be zero, and the solar center $C$ and phase tracking center $O$ are the same. That is precisely what we want as a final result. Since ${\bf R}_{sc}$ is unknown, we actually evaluate the difference between  ${\bf R}_s$ and the corresponding location's theoretical trajectory due to the solar rotation with respect to the phase center $O$ over the period of observation. 

Let the locations of $S$ in the image plane ($x,y$) be denoted as ($x_s,y_s$).  For the time interval of the observation, we have a series of solar images with the calibrated phase containing the position error of the calibrator source. Hence we get a series of reference source positions $ {\bf R}_s^n=(x_s^n,y_s^n)$ where superscript $n$ refers to the discrete time instant $t_n$, $n=1, 2 , ... , N$ during the observation period and $N$ is the total number of the discrete times. The discrete times $t_n$ do not necessarily need to be uniformly distributed over the observation interval. It should be pointed out that, if $S$ represents a stable structure on the sun, its projected position on the solar disk should follow the trajectory due to the solar rotation with respect to the rotation axis crossing the solar center during the observation period. Since we do not know the exact position of solar center, we can nevertheless calculate {$ad ~hoc$} the theoretical trajectories due to the solar rotation with respect to the phase tracking center $O$. The observed positions $ {\bf R}_s^n$ and the calculated theoretical trajectories should coincide to each other if the influence of the calibrator deviation $(l_d,~m_d)$ is eliminated, i.e., the solar center $C$  is coincident with the phase tracking center $O$.  

In reality,  during the elapsed period $ {\bf R}_s^n$ may not always be equal to ${\bf R}_{sc}$ which should follow the correct  trajectory due to the solar rotation with respect to the solar center. We therefore seek to minimize the root mean square (RMS) difference between the observed positions $ {\bf R}_s^n$ and the calculated trajectories with respect to the phase tracking center $O$ as the criterion to judge whether the influence of the calibrator deviation $(l_d,~m_d)$ is eliminated and the solar center $C$ is thus determined. Therefore the  mathematical model for our new position calibration method can be expressed as an optimization problem as follows. Find {$(l_d, m_d)$} such that the RMS value in the following expression reaches a minimum:
\begin{equation}
   \Delta R=\sqrt{\frac{1}{N}\sum_{n=1}^{N} |{\bf R}_s^n-{\bf T}_s({ R}_s^n,R_{\circ})|^2}={\rm min},
  	\label{eq:model}
\end{equation}
where ${\bf T}_{s}({ R}_{s}^n,R_{\circ})$ indicates the above mentioned theoretical trajectory expression of the source $S$. Obviously, it is a function of  ${R}_{s}^n=|{\bf R}_{s}^n|=\sqrt{(x_s^n)^2+(y_s^n)^2}$, and the {radius of the radio sun} $R_{\circ}$. As mentioned above, $N$ is the total number of the discrete time instants during the observation period.

In practice, the optimization problem~(\ref{eq:model}) is solved by an iterative procedure to find unknown $(l_d,~m_d)$ to eliminate the effects of the introduced phase $2\pi(u_{cal}l_d+v_{cal}m_d)$ contained in the initial data for each baseline as described by the following steps. 

{\it Step 0}. Denote the estimated deviation as $(l_a^{(k)},~m_a^{(k)})$ with k=0 representing initial status and the initial values are taken as {$l_a^{(0)}=0$}, and {$m_a^{(0)}=0$}. Set a prescribed but very small positive value $\epsilon$ as the threshold. 

{\it Step 1}. Set $k=k+1$ as an index for the current iteration. For an observational interval (e.g., of several hours), a series of full Sun dirty images can be obtained by equation~(\ref{eq:sunshift}) which contain the influence of the introduced phase  $2\pi(u_{cal}l_d+v_{cal}m_d)$ in the calibrated visibility due to the calibrator deviation  from the phase tracking center. We add a compensation phase term $$-2\pi(u_{cal}l_a^{(k)}+v_{cal}m_a^{(k)})$$ in the corresponding visibility so that the introduced additional phase becomes $$2\pi[u_{cal}(l_d-l_a^{(k)})+v_{cal}(m_d-m_a^{(k)})].$$ A stable radio source on the sun is chosen as the reference position $S$ so that we obtain a set of  $(x_s^n,y_s^n)$ from the dirty images avoiding any solar radio burst interval. {In practice, $l_d, m_d$ are unknown and  inseparable from the corresponding visibility phase terms. Therefore at this stage we actually do not know whether $l_a^{(k)}, m_a^{(k)}$ will compensate the deviation or not}. 

{\it Step 2}. Calculate the theoretical trajectories of the reference position $S$ in the same way as if the reference source $S$ rotates on the solar disk with respect to the phase tracking center $O$ in Fig.~\ref{deviation}.

{\it Step 3.} Evaluate the objective function or the RMS value $\Delta R$ of the difference between the observed positions and the theoretical trajectory positions of the reference source for the observational period as described in the optimization model~(\ref{eq:model}). If the RMS value $\Delta R$ does not decrease and the difference between consecutive trial deviations $$\sqrt{|l_a^{(k)}-l_a^{(k-1)}|^2+|m_a^{(k)}-m_a^{(k-1)}|^2}\geq \epsilon,$$ we modify $(l_a^{(k)},m_a^{(k)})$  to further reduce the influence of the additional phase term $2\pi(u_{cal}l_d+v_{cal}m_d)$ due to the calibrator deviation and repeat from {\it Step 1}. Otherwise continue to the next step.

{\it Step 4.}  If the difference between consecutive iterations is less than $\epsilon$ and the RMS value $\Delta R$ does not increase, or  the convergence criterion has been satisfied,  the RMS value $\Delta R$ has reached its minimum value. Therefore the unknown deviation  ($l_d, m_d$) has been inferred to be  ($l_a^{(k)}, m_a^{(k)}$) approximately. Hence the influence due to calibrator deviation from the phase tracking center has been largely eliminated and the iterative procedure can be terminated

Optimization techniques can be incorporated in the above procedure. We can also rotate all time series images to the image location corresponding to a fixed instant in order to evaluate the distribution of the reference source during the observation period. Obviously the reference source positions at different instants should converge to the same place on the image corresponding to the certain instant if it is a stable source. 

 The proposed phase calibration process will be demonstrated in the following section for its correctness and merits in removing calibrator position discrepancies.

\section{Application to Eliminate Calibrator Position Deviation}
\label{section4}

{We have conducted several simulation case studies and realistic data processing for MUSER observations to validate the applications of the proposed new method}.

\subsection{Simulation}
\label{subsect4.1}

We first perform a simulation experiment. The model consists of a uniform disk and two Gaussian sources with different widths in their intensity distribution. {In the first image of this simulation test, the source on the left at ${1}/{2} R_{\circ}$ at a position angle 154$^{\circ}$ from the x-axis is chosen as the reference source, and its peak intensity is 20 times that of the background. The "date" of the simulation data is taken to be November 22, 2015, and the observation period is from 02:05 UT to 06:05 UT. On that particular day, the solar $P$ angle was around 19$^{\circ}$. For MUSER, the local meridian time is roughly 4 UT. The working frequency is 1.7125 GHz}.

\begin{figure}
 \centering
                         \includegraphics[width=0.31\textwidth]{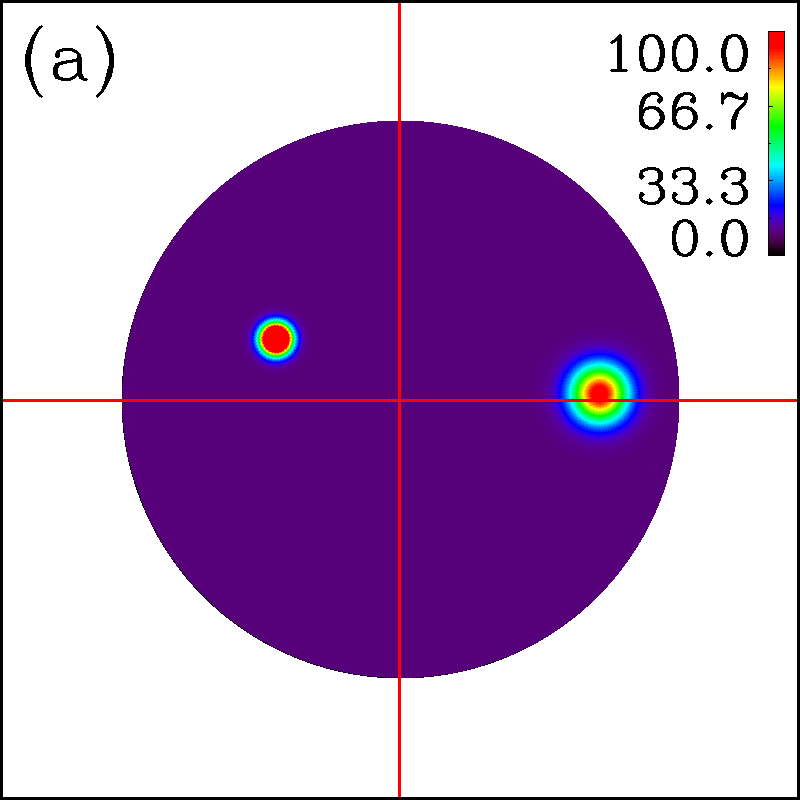}
			\includegraphics[width=0.31\textwidth]{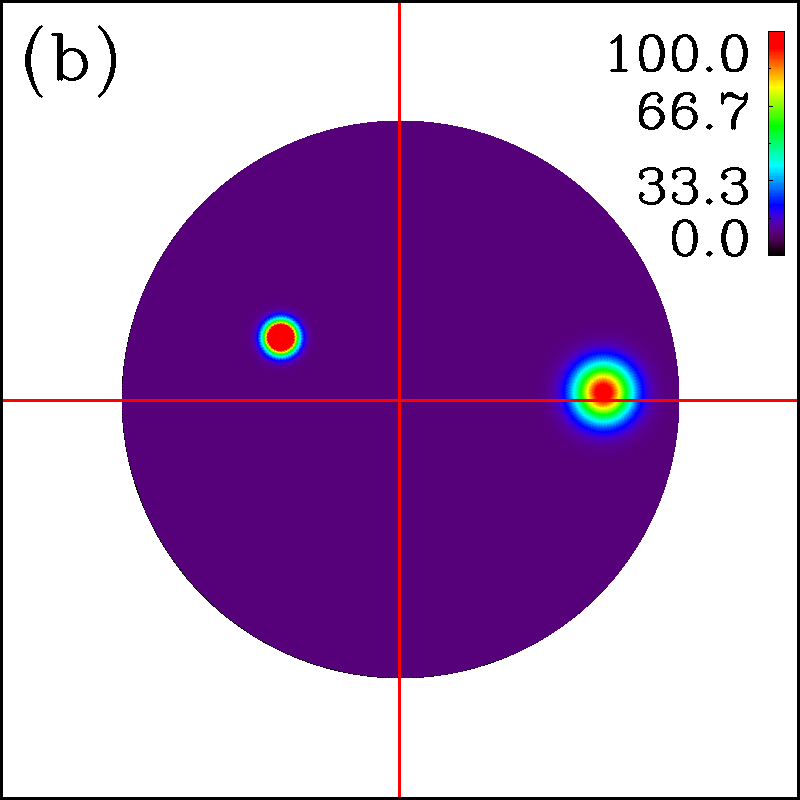}
			\includegraphics[width=0.31\textwidth]{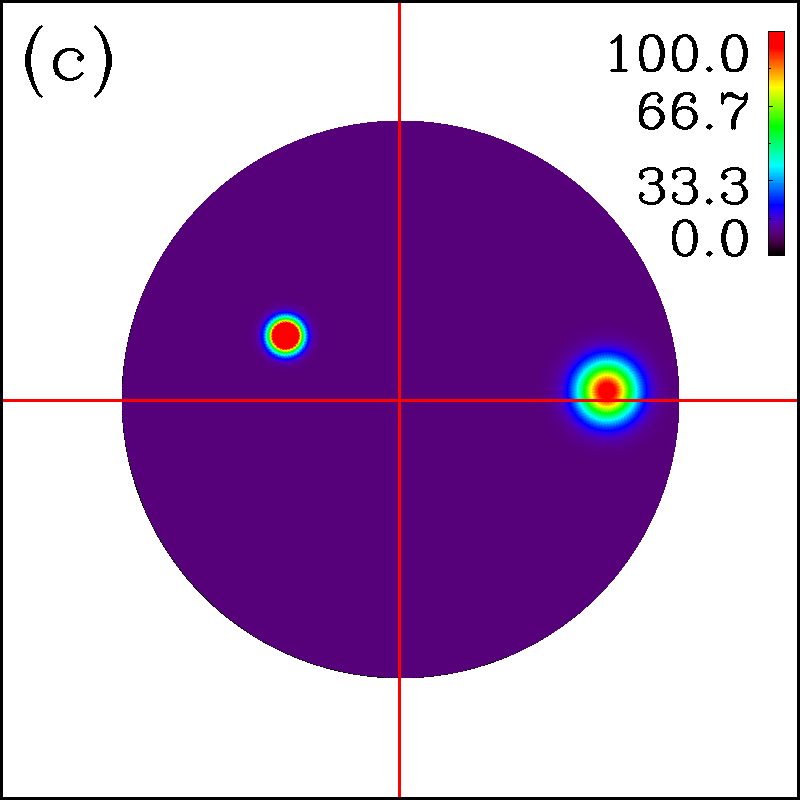}\\
			\includegraphics[width=0.31\textwidth]{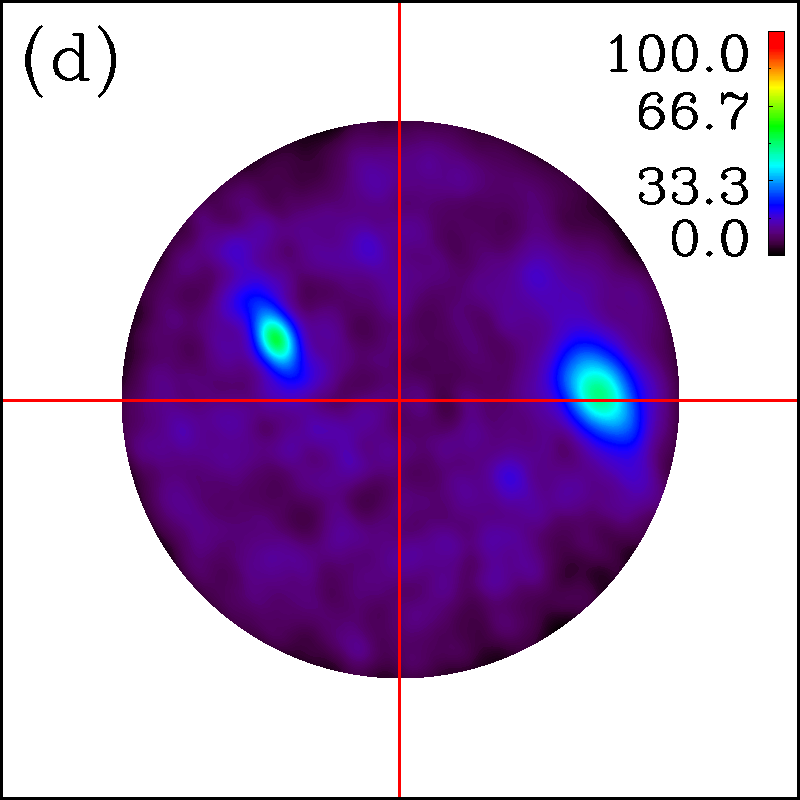}
			\includegraphics[width=0.31\textwidth]{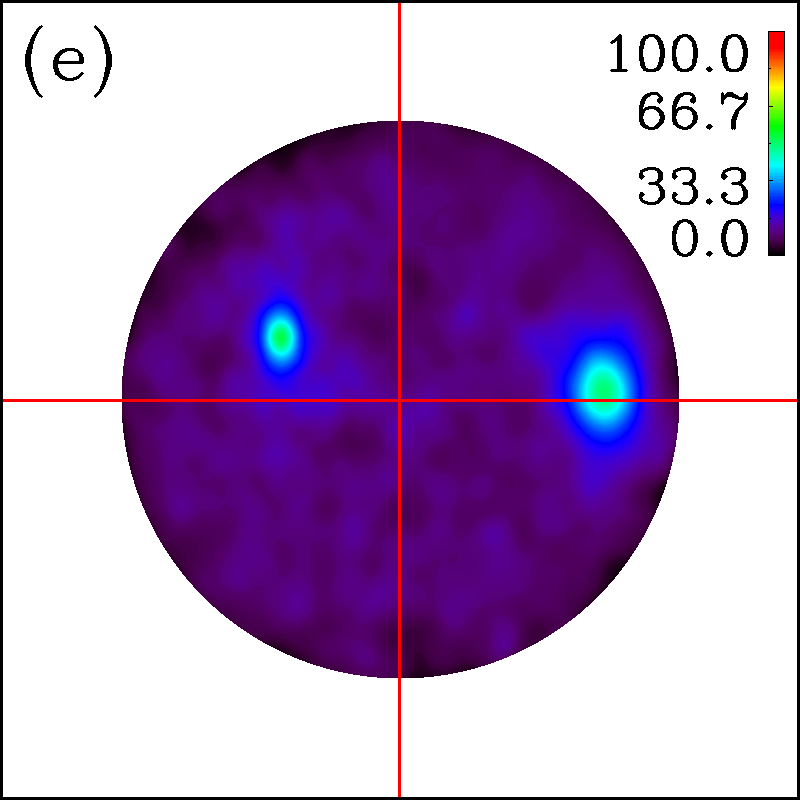}
			\includegraphics[width=0.31\textwidth]{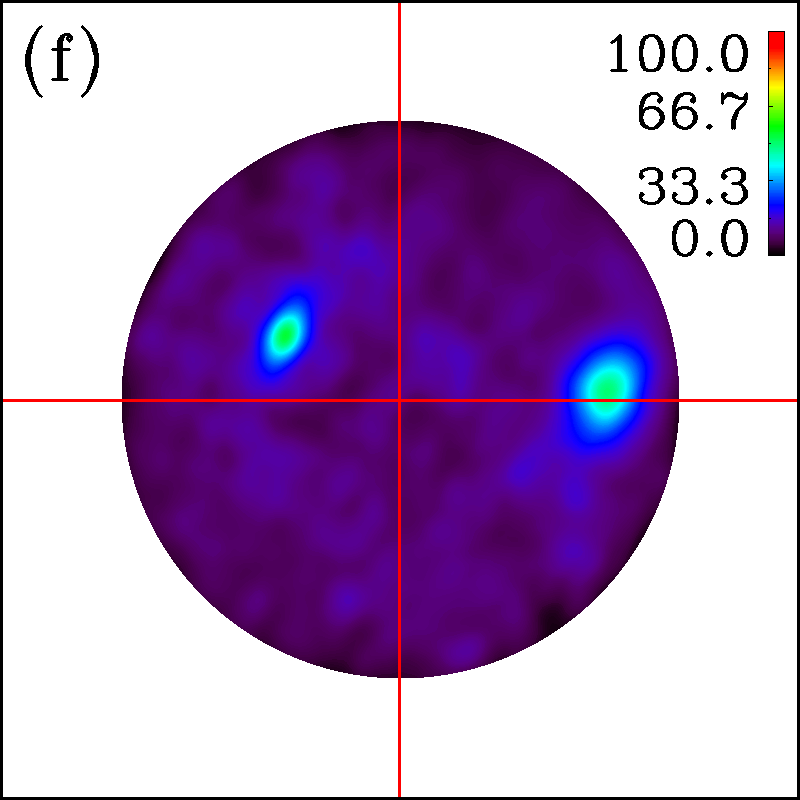}\\


\includegraphics[width=0.31\textwidth]{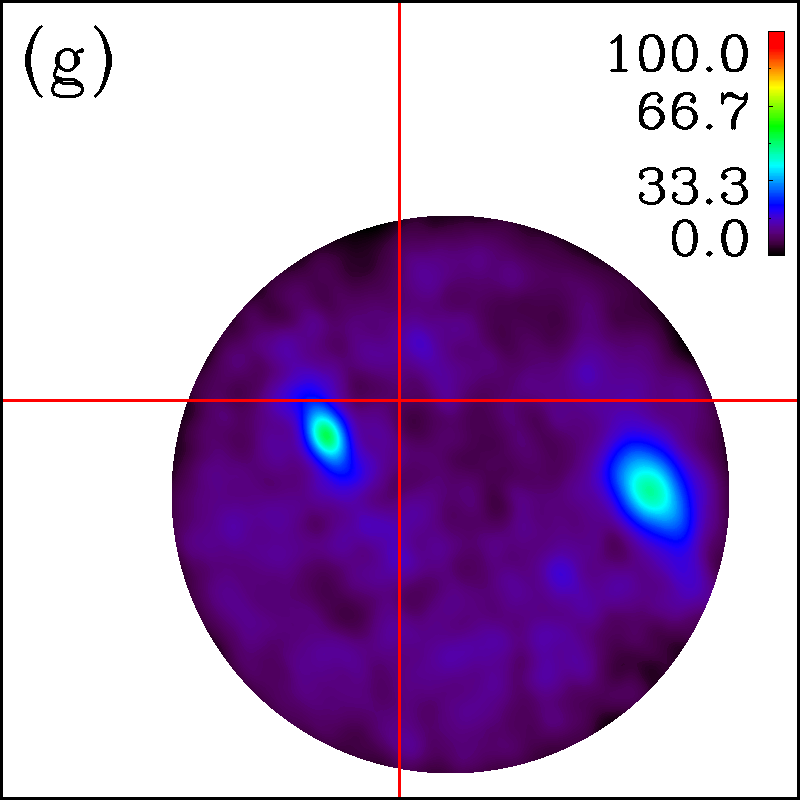}
\includegraphics[width=0.31\textwidth]{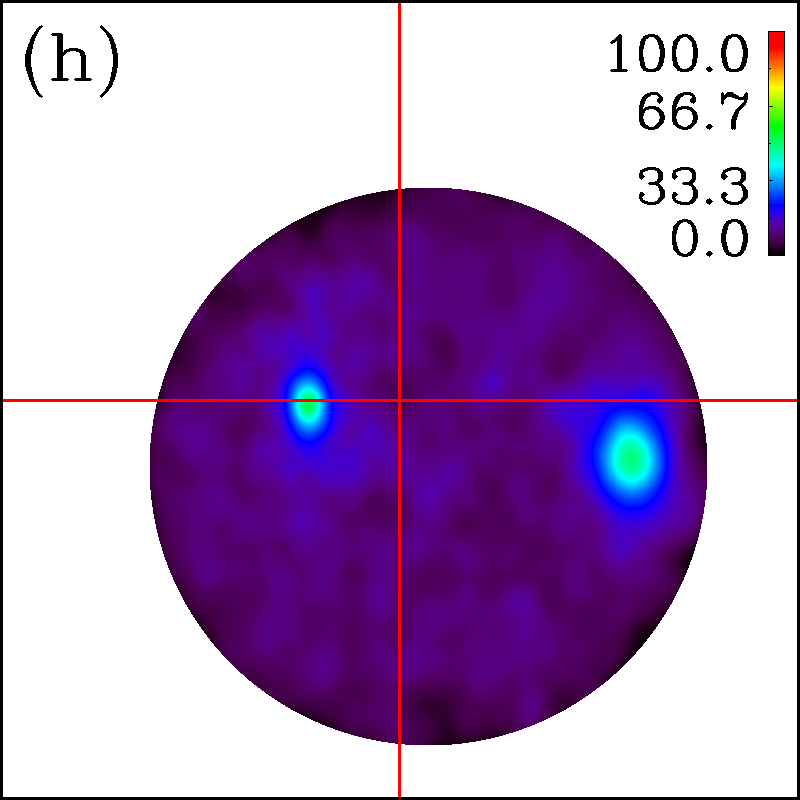}
\includegraphics[width=0.31\textwidth]{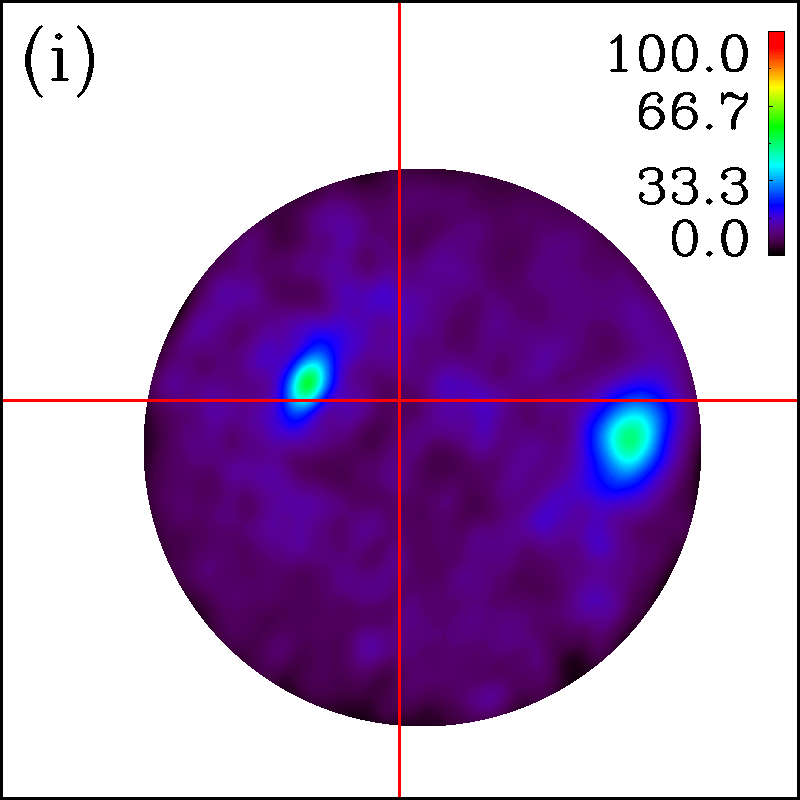}
		 \caption{{\it Upper row}: the simulation model at (a) 02:05:00, (b) 04:05:00, and (c) 06:05:00 UT respectively.  {\it Middle row}: (d-f) the corresponding dirty images of the model if they were observed by MUSER. {\it Bottom row}:  (g-i) the corresponding dirty images further affected by the calibrator position offset.}
    \label{simulation1}
\end{figure}

 The first row of solar model images in Fig.~\ref{simulation1} {is thus established. From left to right, the three images are assumed to correspond to  (a) 02:05:00, (b) 04:05:00, and (c) 06:05:00 UT. The solar rotation has been taken into account, which causes the position of the radio source to vary over the  three different instants. If MUSER-I were to observe the aforementioned solar model images, we would obtain the dirty maps shown in the middle row of} Fig.~\ref{simulation1}(d-f). 

A set of phases $2\pi(u_{cal}l_d+v_{cal}m_d)$ with the initial deviation of $l_d=-1.4^{\prime}, m_d=4^{\prime}$ are then added to the phases of the visibilities corresponding to the model dirty image  at different time instants. {The deviation values are significantly larger than the angular resolution of $0.245^{\prime}$ at the 1.7125 GHz MUSER observing frequency, emulating the practical situation of introducing a set of additional phases in phase calibration due to the calibrator position error deviating from the phase tracking center.} The three images in the bottom row in Fig.~\ref{simulation1}(g-i) are respective dirty images affected by the calibrator position deviation and they are the initial data for our analyses. 

\begin{figure}
  \centering
			\includegraphics[width=0.95\textwidth]{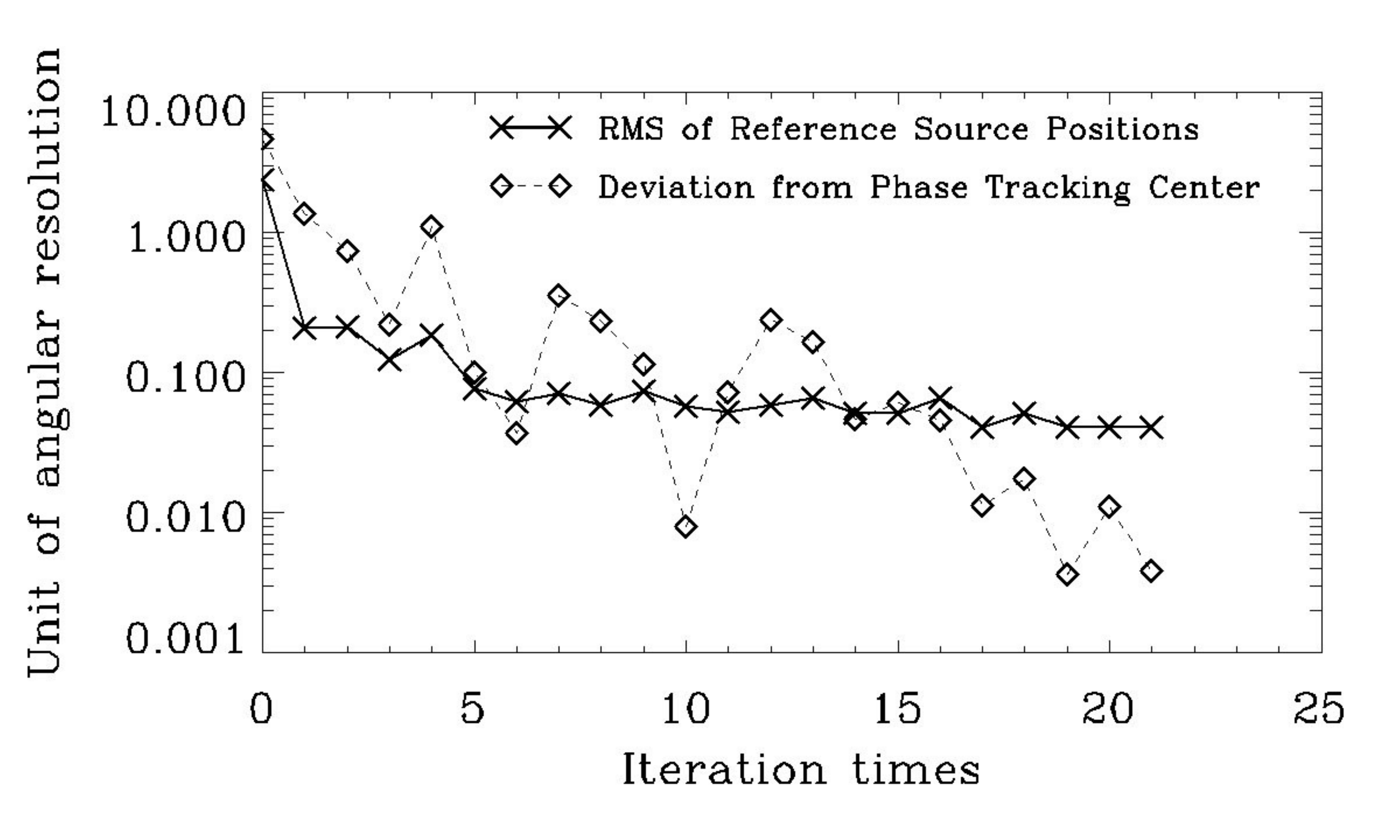}
		 \caption{ The iteration history  {\it a posteriori} of the RMS values ($\Delta R$ of the reference source positions (solid line and "cross" symbol) and the deviation from the phase tracking center (dotted line and "diamond" symbol) for the solar model simulation. These values are expressed in unit of the angular resolution at the MUSER observing frequency.  }
    \label{iteration}
\end{figure}

We can readily apply the iterative algorithm as described in section~\ref{subsect3.2} to solve the mathematical model~(\ref{eq:model}) for the simulated data. Fig.~\ref{iteration} shows the {\it a posteriori} iteration history of the RMS error $\Delta R$ of the reference source positions as expressed in model~(\ref{eq:model}). It can be seen that the RMS error $\Delta R$ decreases to a small fraction of the angular resolution at MUSER observing frequency after about ten iterations. The image deviation  {\it a posteriori} from the phase tracking center also converges to the expected value after a few iterations. 

\begin{figure}
  \centering
			\includegraphics[width=0.31\textwidth]{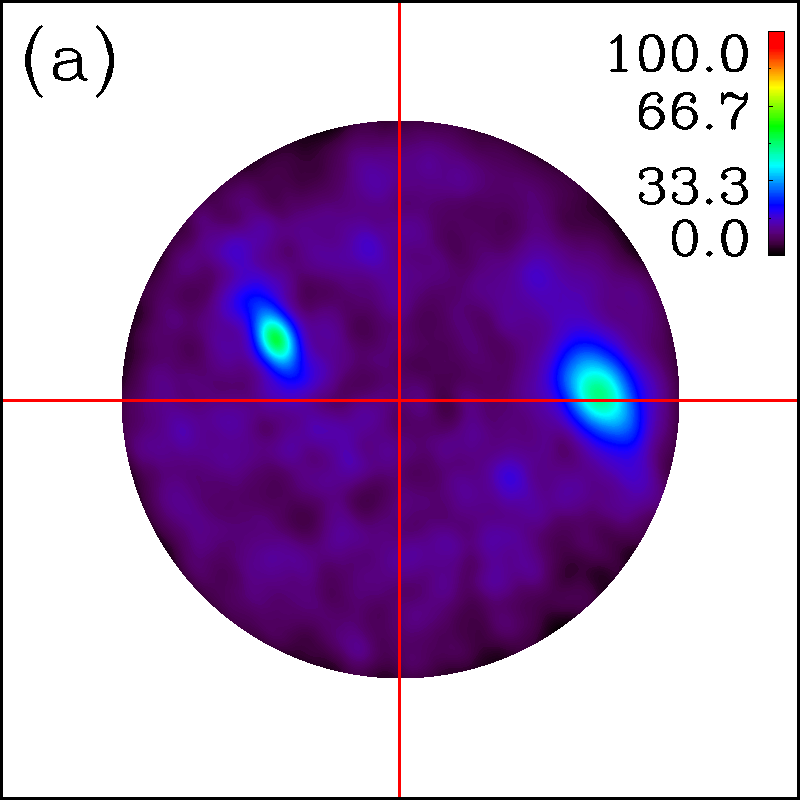}
			\includegraphics[width=0.31\textwidth]{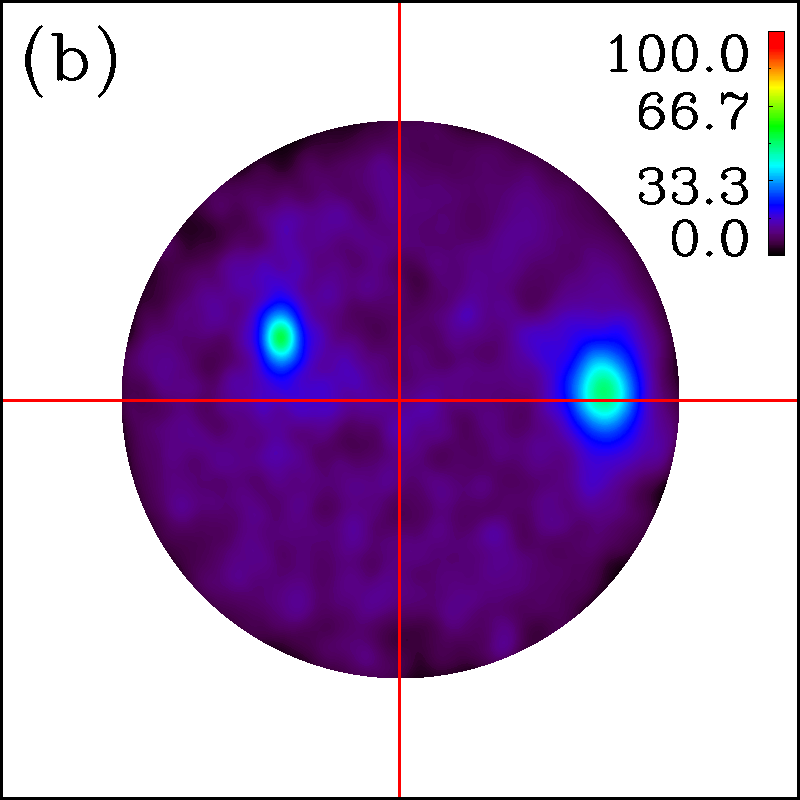}
			\includegraphics[width=0.31\textwidth]{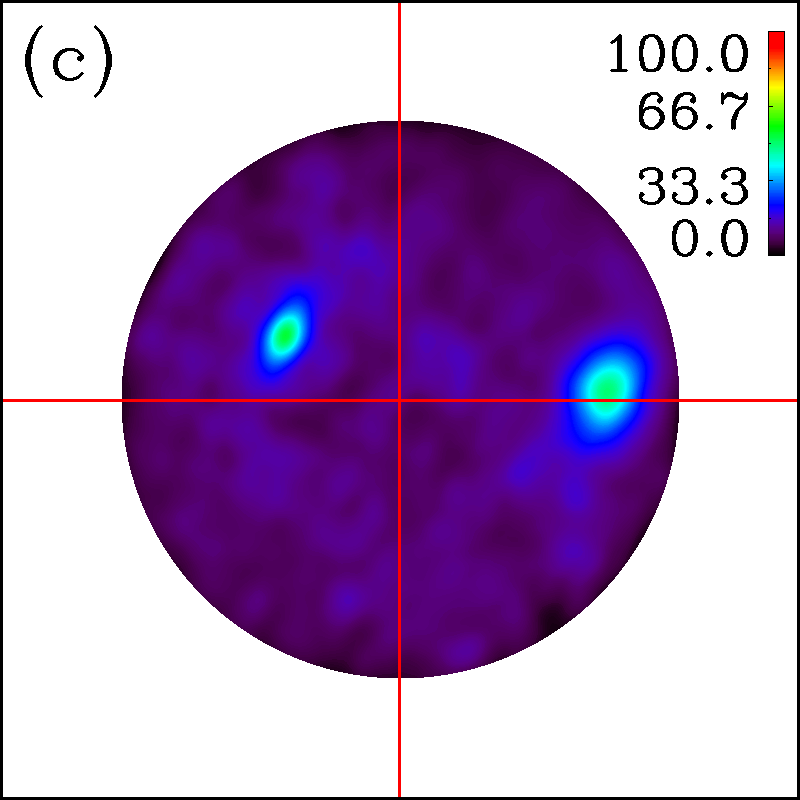}\\
			\includegraphics[width=0.31\textwidth]{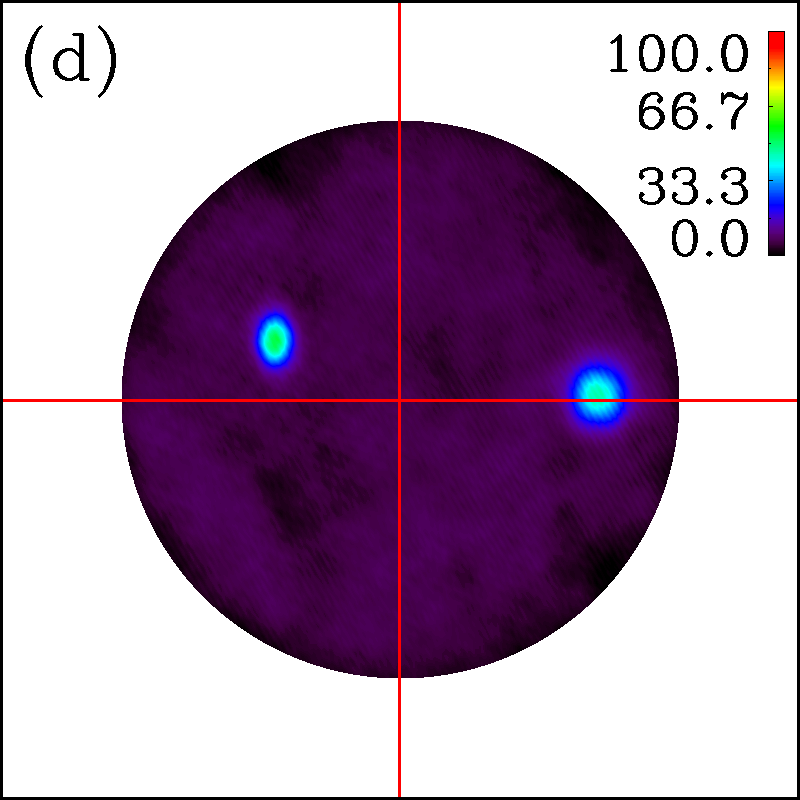}
			\includegraphics[width=0.31\textwidth]{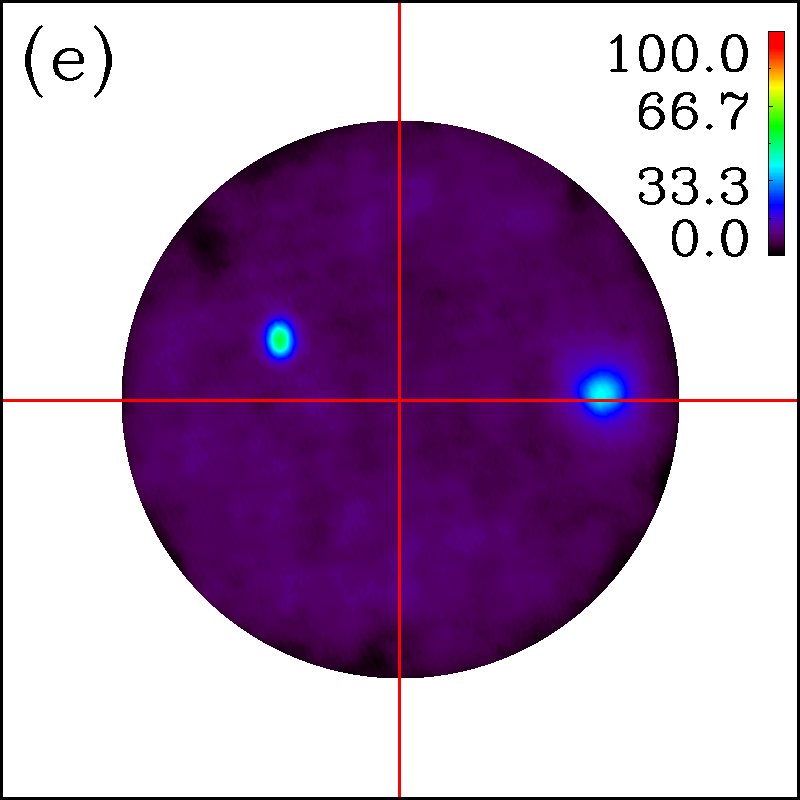}
			\includegraphics[width=0.31\textwidth]{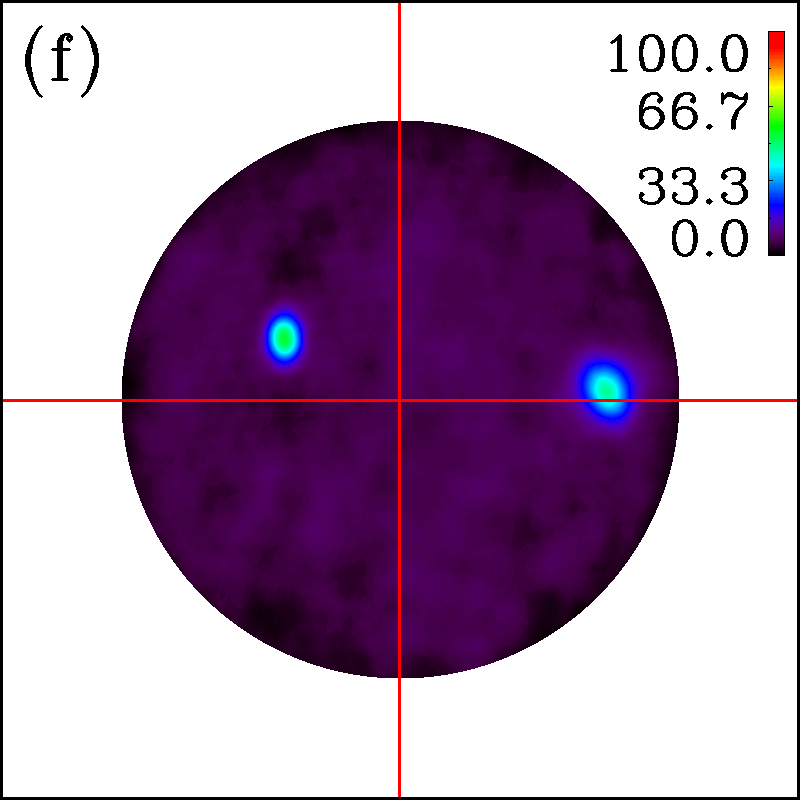}
		 \caption{ {\it Upper panels}: The position calibrated dirty images of the model at (a) 02:05:00, (b) 04:05:00, and (c) 06:05:00 UT respectively.   {\it Bottom panels}: (d-f) The corresponding cleaned images of the model. }
    \label{simulres}
\end{figure}

The position calibrated dirty images and restored clean images are shown in Fig.~\ref{simulres}. These results are satisfactory as compared with the original model, as seen in the top two rows of Fig.~\ref{simulation1}.  It should be noted that all the restored radio images were obtained through Hogbom CLEAN algorithm \citep{Thompson+2017} unless stated otherwise. 

 \begin{figure}
 \centering
			\includegraphics[width=0.32\textwidth]{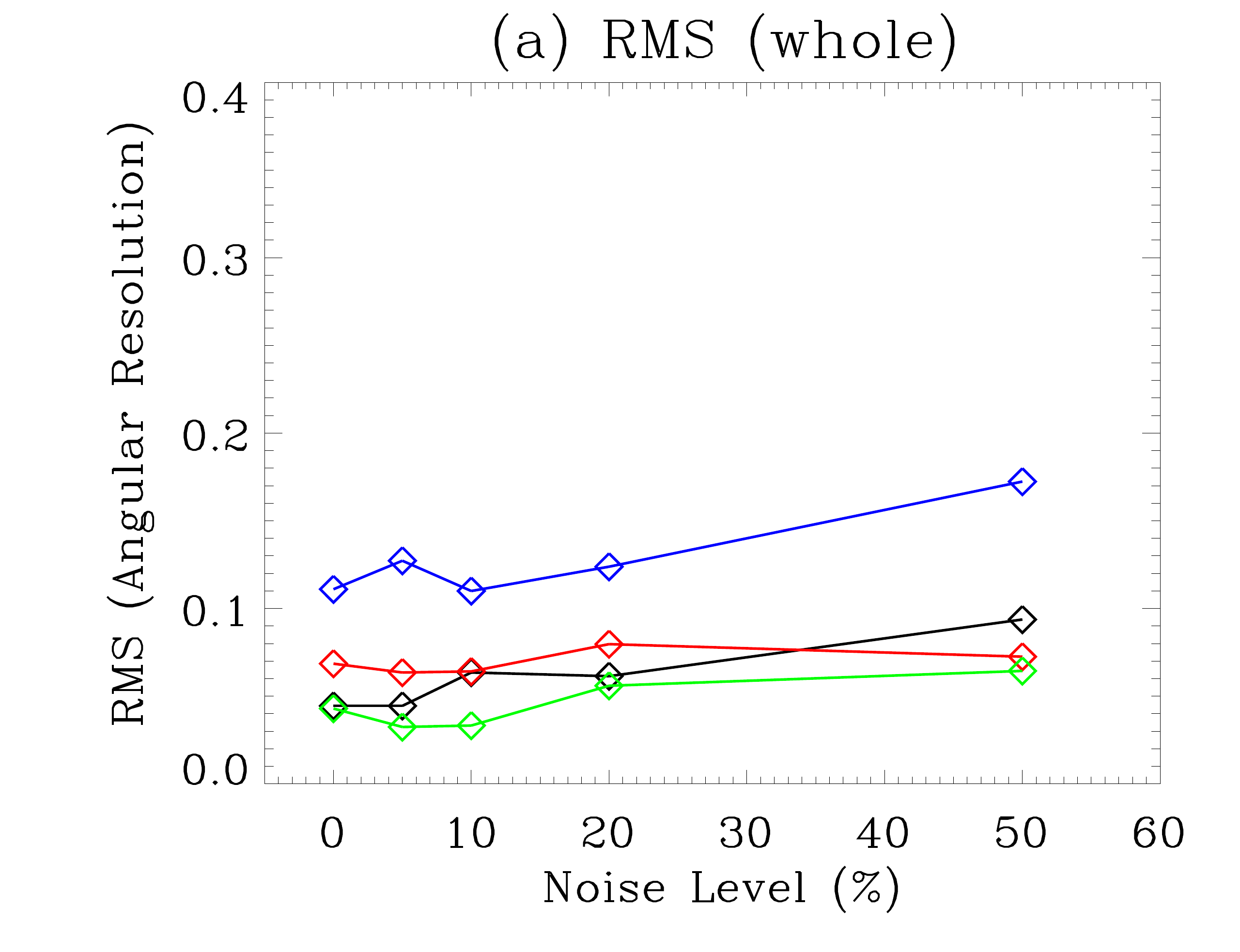}
			\includegraphics[width=0.32\textwidth]{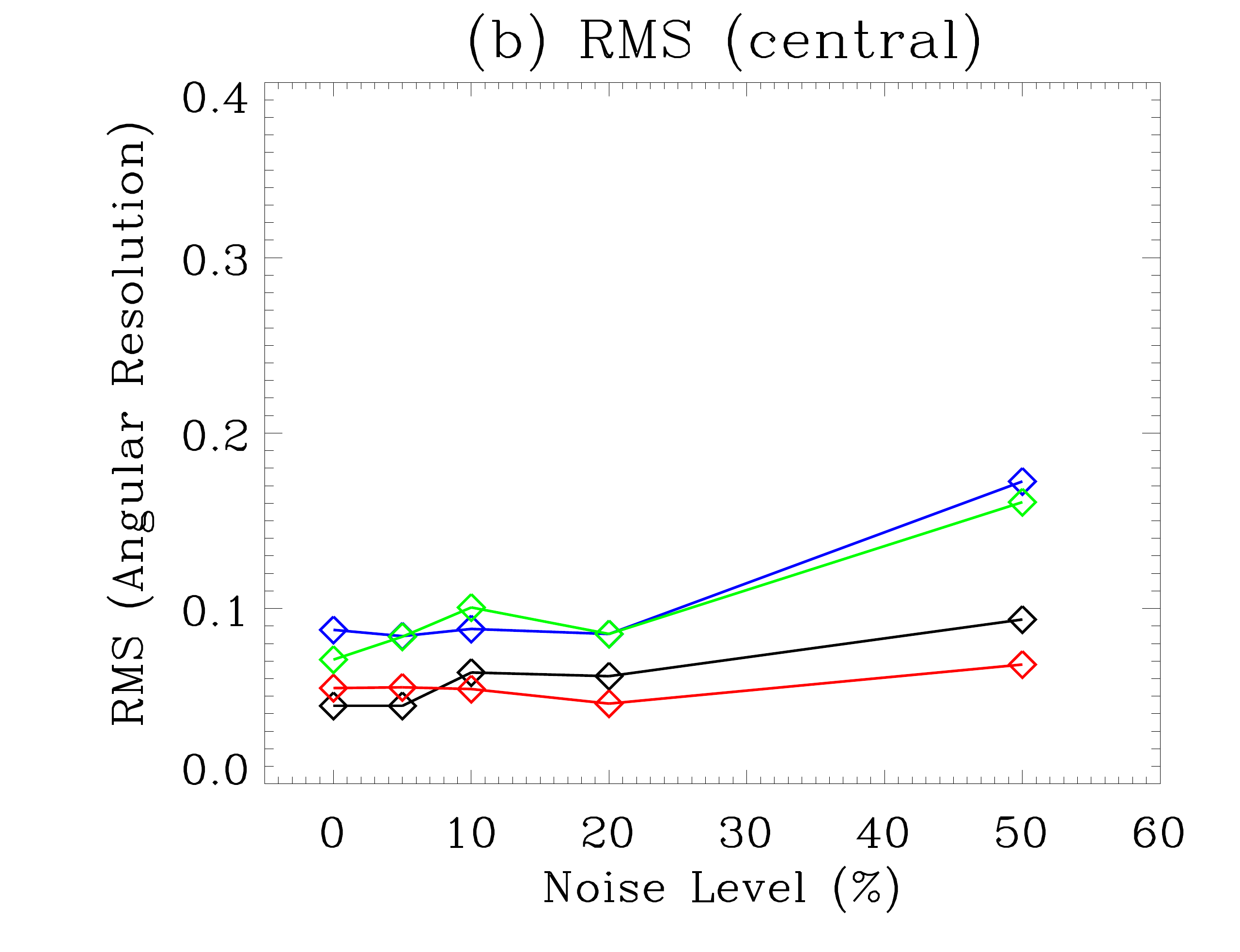}
			\includegraphics[width=0.32\textwidth]{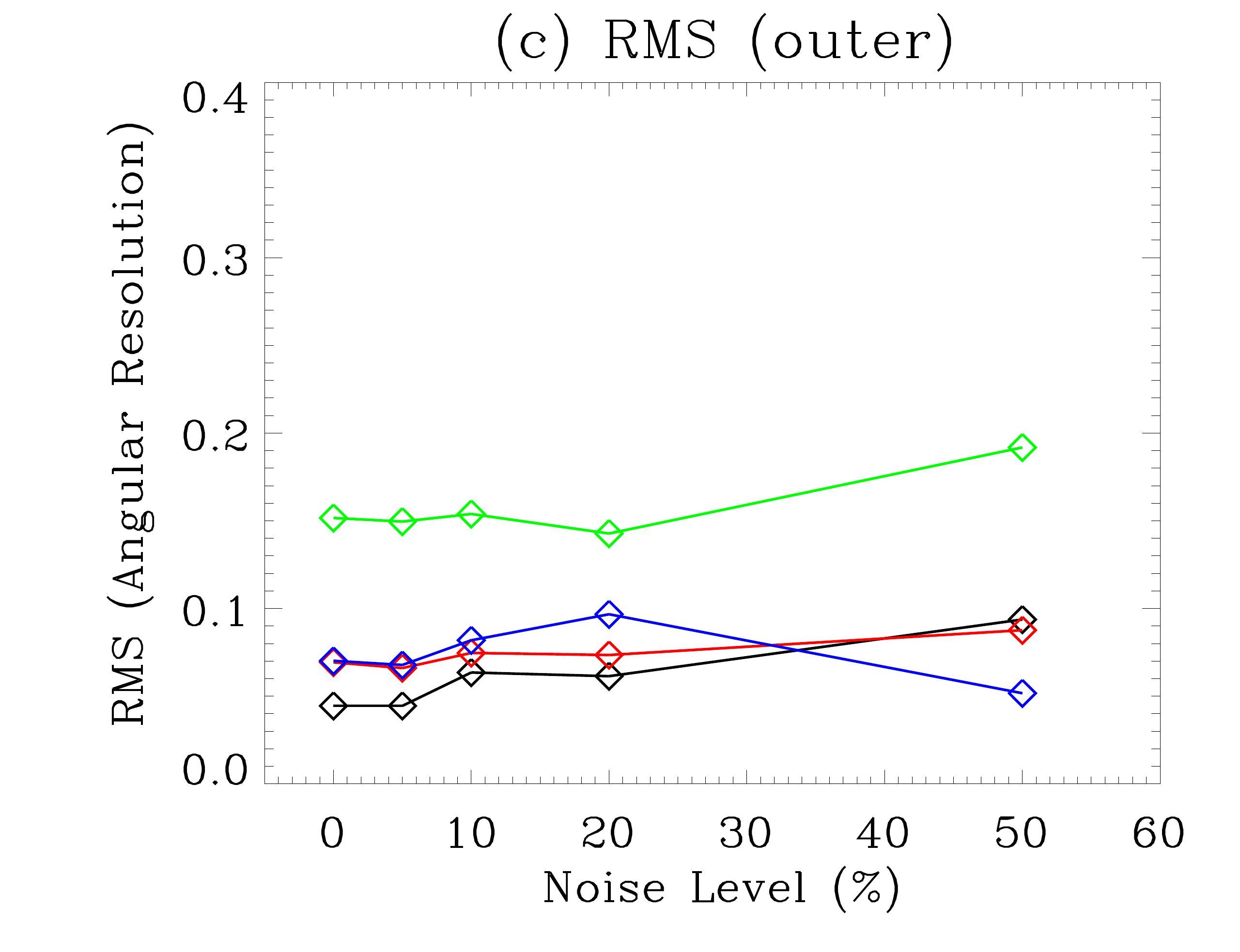}\\
\hspace{0.02em}
			\includegraphics[width=0.32\textwidth]{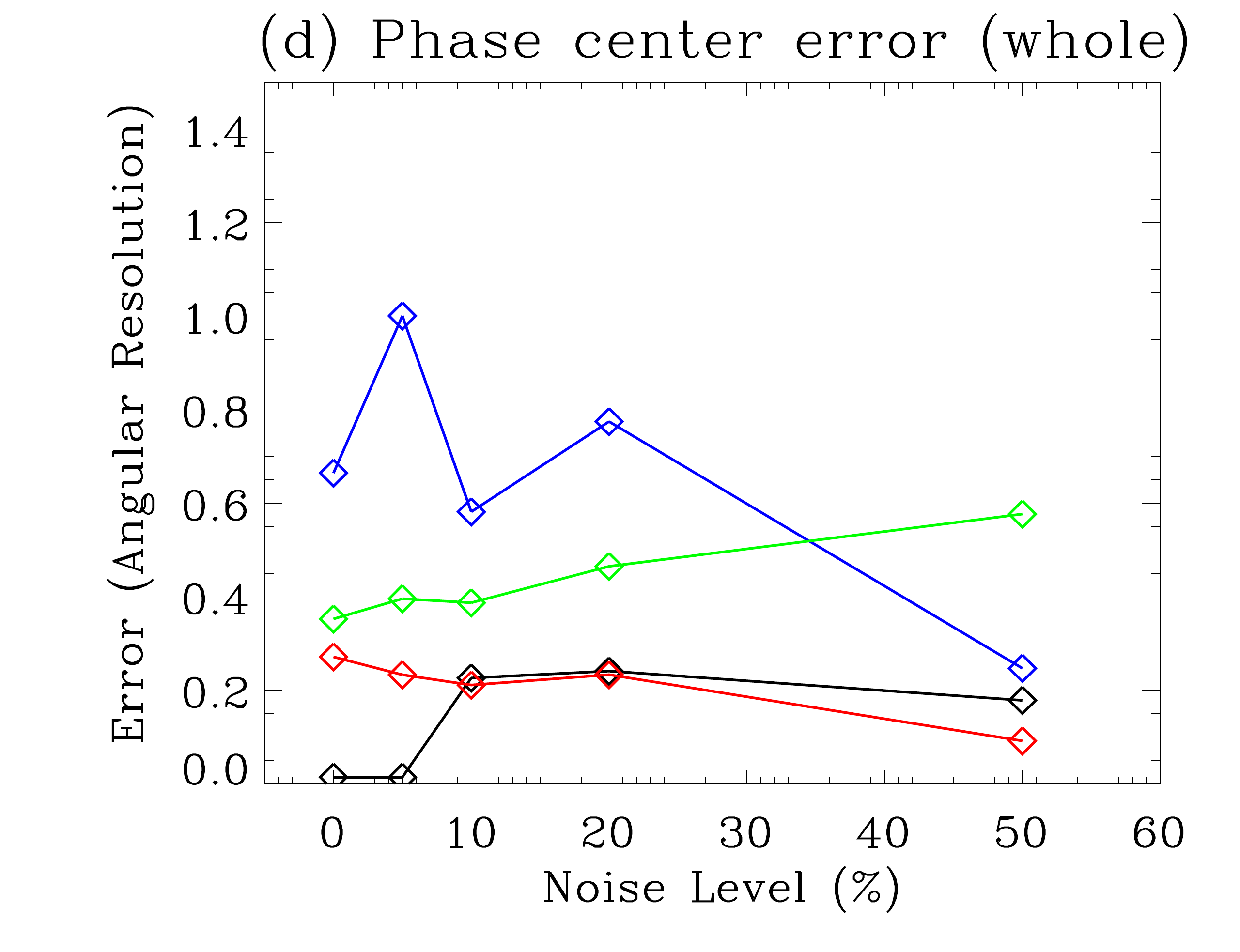}
			\includegraphics[width=0.32\textwidth]{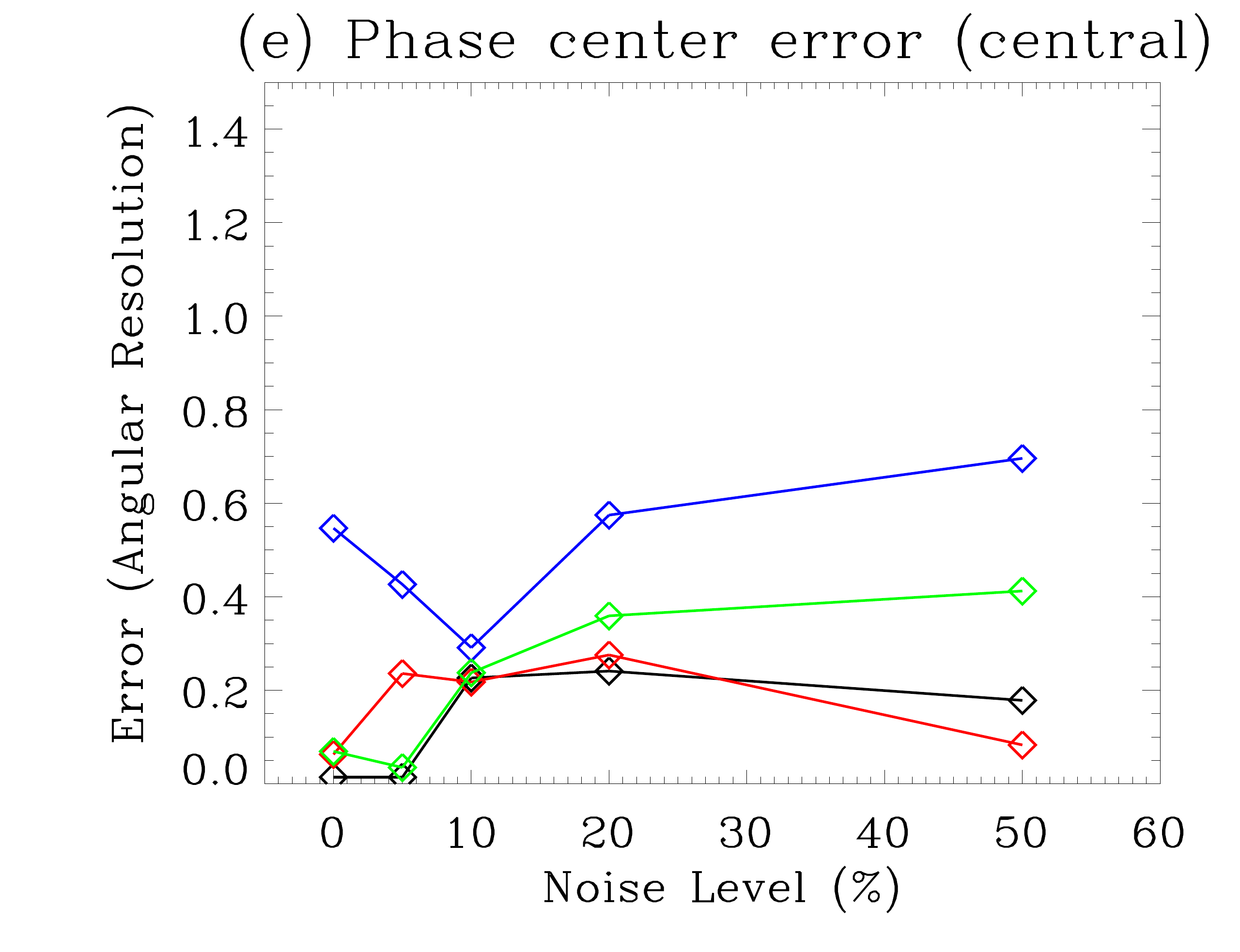}
			\includegraphics[width=0.32\textwidth]{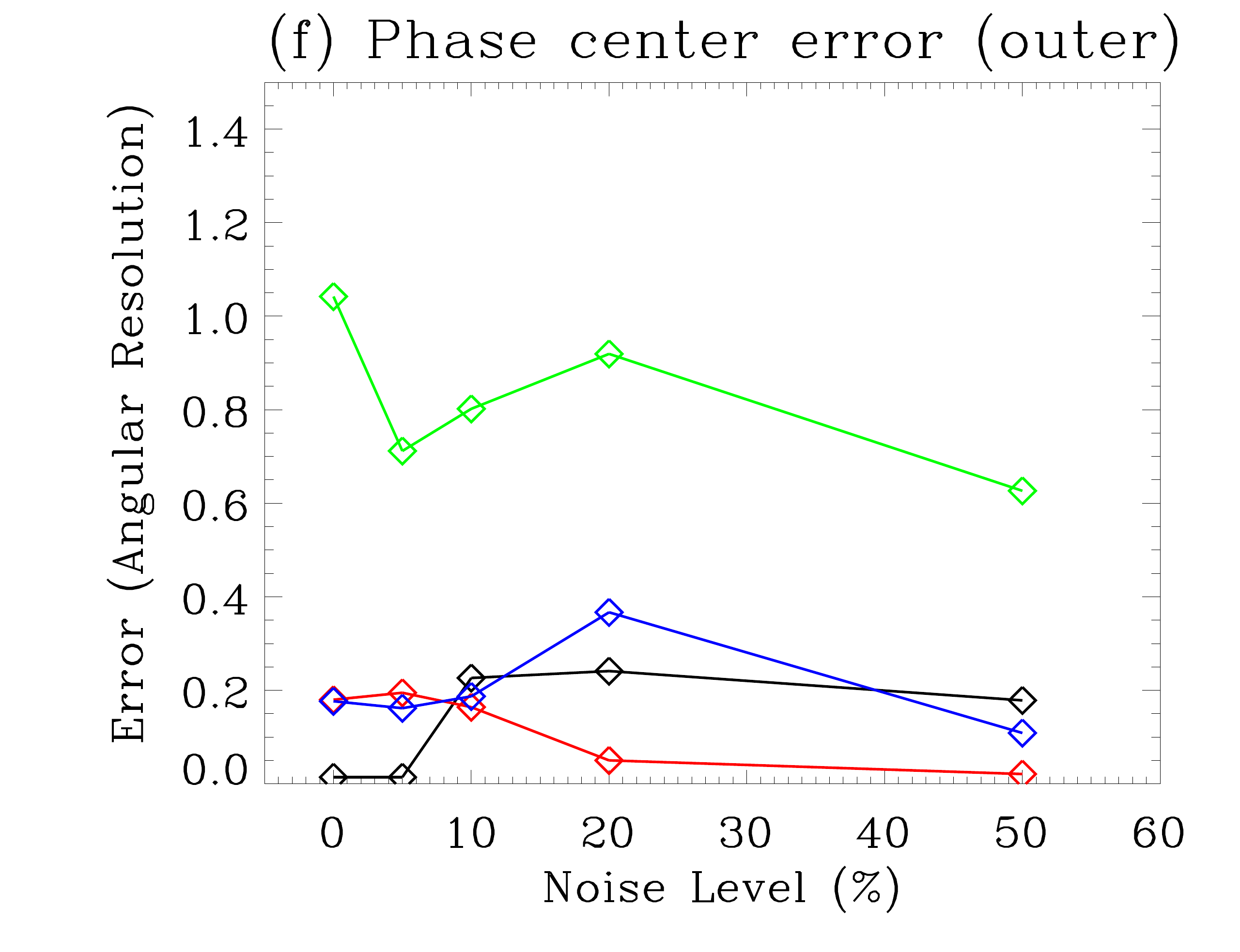}
 \caption{After convergence of the model simulation, the distributions of the objective function RMS $\Delta R$ versus the noise levels added in fraction of the quiet sun intensity under three situations: (a) with some or no antennas flagged; (b) with some or no antennas flagged but confined in the central area only; (c) with some or no antennas flagged but in the outer area only. The corresponding distributions of the recovered phase center error versus the noise levels added in fraction of the quiet sun intensity under similar situations are shown in (d-f). The black, red, blue, or green color lines indicate the cases when none (0\%), 10\%, 30\% or 50\% of total antennas is flagged.}
    \label{sim}
\end{figure}

 To imitate real-world scenarios, we introduced varying amounts of noise to the simulation cases, such as 5\%,10\%,20\%, and 50\% quiet sun intensity, respectively, as random normal distributed noises. The situation in which not all antenna baselines are functioning has also been considered. Normally these baselines are flagged so that they are not involved in the aperture synthesis imaging process. We looked at three situations in which the flagged baselines are confined to short, long, or all baselines. Here, the baseline lengths $<200$ m are regarded as the short; otherwise it is regarded as long. For the sake of convenience, we simply flag antennas and consider the central antennas in the inserted panel of Fig.~\ref{antennas position} {to be short-baseline antennas, even though they may be long-baseline antennas with other antennas outside the central area. The number of central area antennas in the MUSER-I array, as shown in} Fig.~\ref{antennas position}, is 19 and the number of rest antennas is 21. As a result, in the actual simulations, we simply flag the fractions of 10\%, 30\% and 50\% of total antennas that are confined to (i) only the core area, (ii) only the outside area, and (iii) all antennas. In this way, we may evaluate the practicality of the proposed position calibration method.

In general the problem converges after about twenty iterations in all cases. Fig.~\ref{sim} {shows the {$a~ posteriori$} results of both the objective function RMS error ($\Delta R$) of the reference source positions and the recovered phase centre error versus different noise levels ranging from noise-free, 5\%, 10\%, 20\% to 50\% quiet Sun intensity, and varied flagging antenna fractions ranging from 0\%, 10\%, 30\% to 50\% total antennas but limited either in (i) the central area only, (ii) the outside area only, or (iii) with no restrictions.} From the simulation results it can be seen that in most  cases the recovered position errors are with a small fraction of the corresponding spatial angular resolutions with available baselines. Only in a few cases the calibration position error may reach around one-fold spatial angular resolution.  It should be noted that in those cases many antennas were flagged out in the exterior of the central area, i.e., most long baselines were flagged. Because longer baselines lead to higher angular resolution, the majority of information corresponding to increased angular resolution is lost. Nevertheless the recovered position errors with different noise levels are still around one-fold angular resolution even under these circumstances. The above results indicate that the proposed new position calibration method is valid and practical.

\subsection{Calibration of MUSER Observational data}
\label{subsect4.2}

\begin{figure}
  \centering
 	\includegraphics[width=0.95\textwidth]{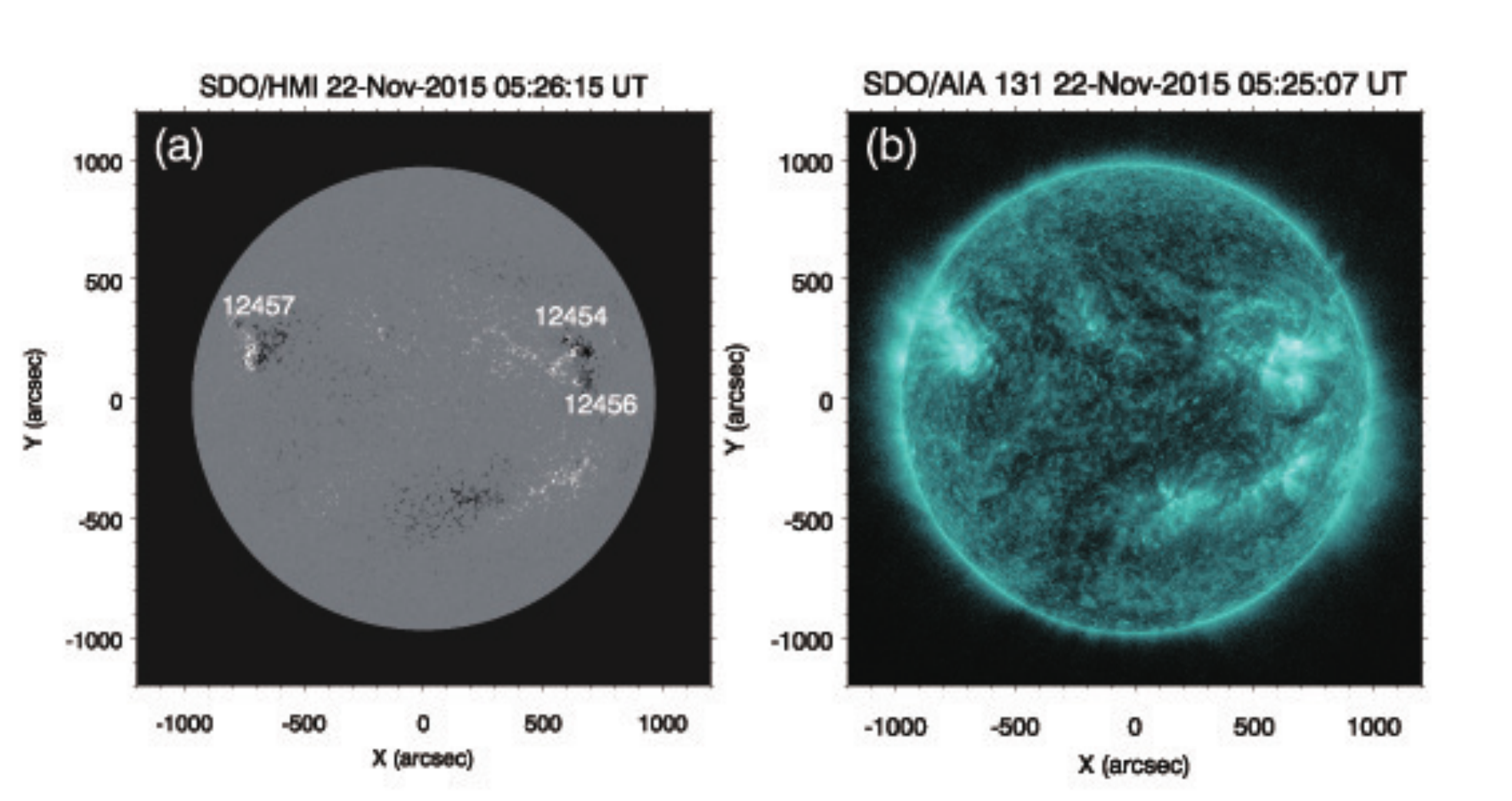}
   \caption{Observations of the sun on November 22, 2015: (a) SDO/HMI magnetogram at 05:26:15 UT, (b) SDO/AIA 131 \AA ~EUV image at 05:25:07 UT.}
    \label{event1}
\end{figure}

 To investigate the validation of this method using actual observational data, we chose November 22, 2015 as the date. The SDO/HMI magnetogram and AIA 131\AA ~EUV image are shown in Fig.~\ref{event1}. It can be seen that there were several solar active regions on that day,  AR12457 N11E36 (-562$^{\prime\prime},159^{\prime\prime}$), AR12456  N06W55 ($793^{\prime\prime},82^{\prime\prime}$), and AR12454 N13W53(757$^{\prime\prime},199^{\prime\prime}$). There was a GOES SXR C5.6 class flare in AR12454 starting at 05:31 UT, peaking at 05:38 UT and ending at 05:41 UT. The radio source observed by MUSER-I at AR12457 was chosen as the reference source because it was more stable over the observational period.

In general, we use the radio signal of the Geosynchronous weather satellite operating at around 1.7GHz as a calibrator for MUSER data calibration \citep{Wang+2019}.  This satellite has another working frequency around 1.68 GHz.  The difference in the angular resolution of the MUSER-I array between the two frequencies is less than 1.5 percent. Hence we apply both frequencies for the phase calibration. Since they are both from the same calibrator position, the location differences of any sources in the target images from both frequencies should measure the original position variations corresponding to different frequencies, which was with a RMS value $\Delta R$ of 0.727 in units of the angular resolution before the phase calibration. The present method is then used to calibrate the solar radio images at these two frequencies. The solar radio model in \citep{Tan+2015} has been employed to consider the factor of the frequency-dependent radii. 

\begin{figure}
  \centering
  \includegraphics[width=0.95\textwidth]{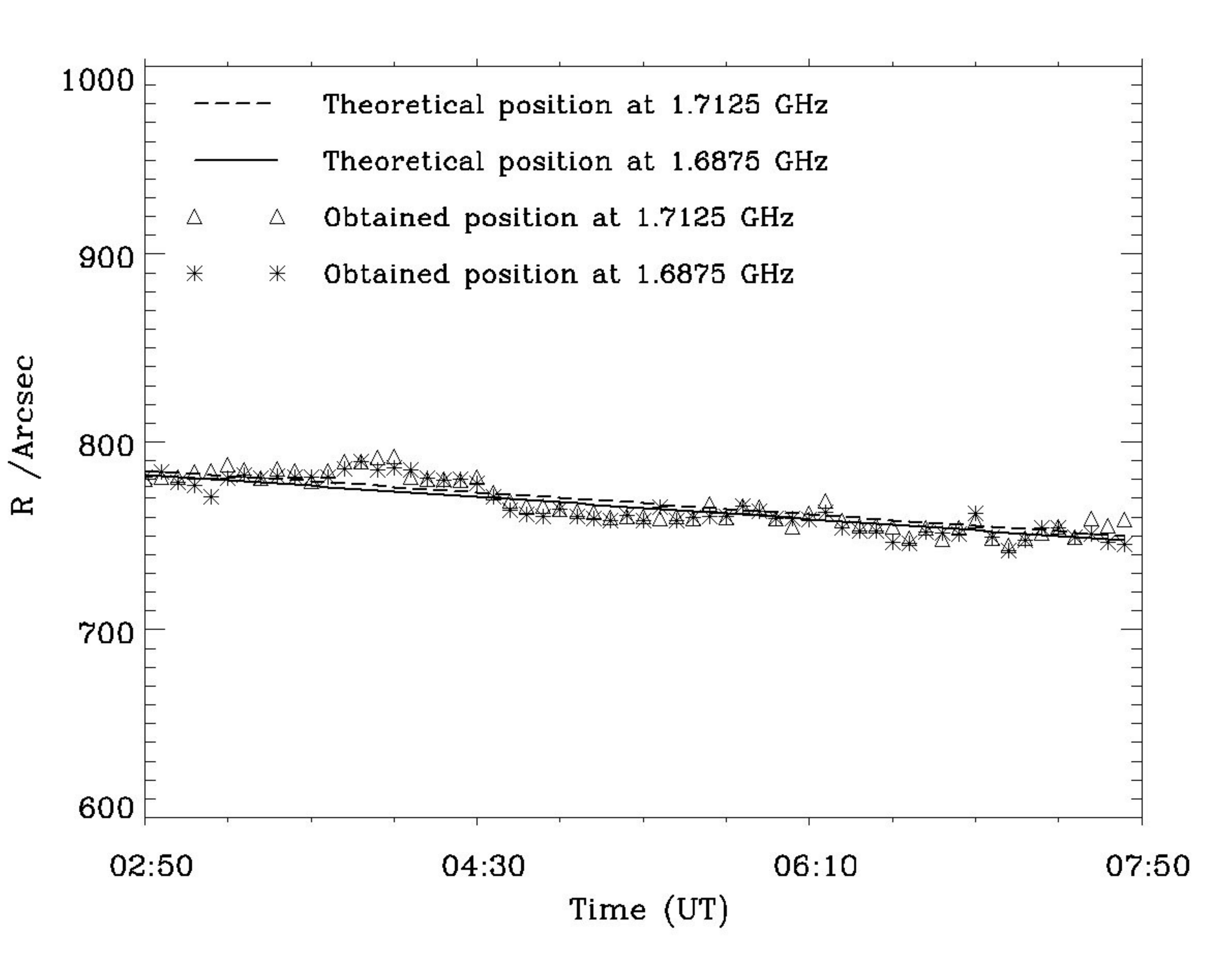}
  \caption{Position tracking of the reference radio sources at 1.7125 GHz (marked with triangle)  and 1.6875 GHz (marked with stars) observed by MUSER-I on November 22, 2015. The abscissa is universal time and the ordinate is R, the distance from the reference source to the solar center  in arc second. The dashed  and solid lines indicate the theoretic trajectory results of tracked positions at 1.7125 GHz and 1.6875 GHz, respectively.  }
  \label{track}
\end{figure}

 The reference radio source at different times is tracked, expressed in terms of the distance from its
location relative to the solar center, or its radius on the solar disk, $R$, and displayed in Fig.~{\ref{track}} with triangle and star symbols denoting 1.7125 GHz and 1.6875 GHz, respectively.  The RMS value $\Delta R$ of the location differences of this reference source for both frequencies is slightly changed as 0.696 in unit of the angular resolution after the calibration. The dashed and solid lines in Fig.~{\ref{track}} show the corresponding trajectories of the reference radio source based on the theoretical calculation described in the previous section. 

\begin{figure}
  \centering
  \includegraphics[width=0.45\textwidth]{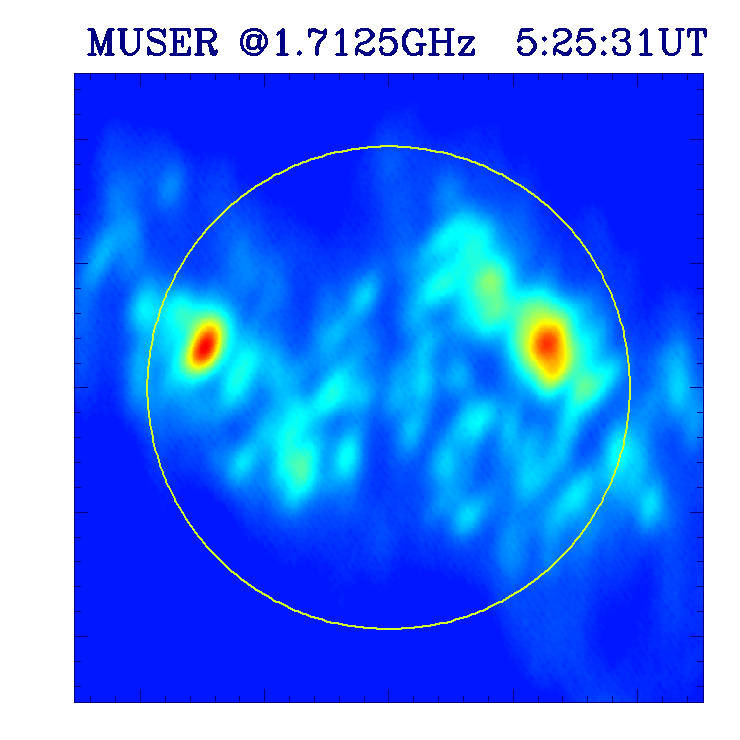}
  \includegraphics[width=0.45\textwidth]{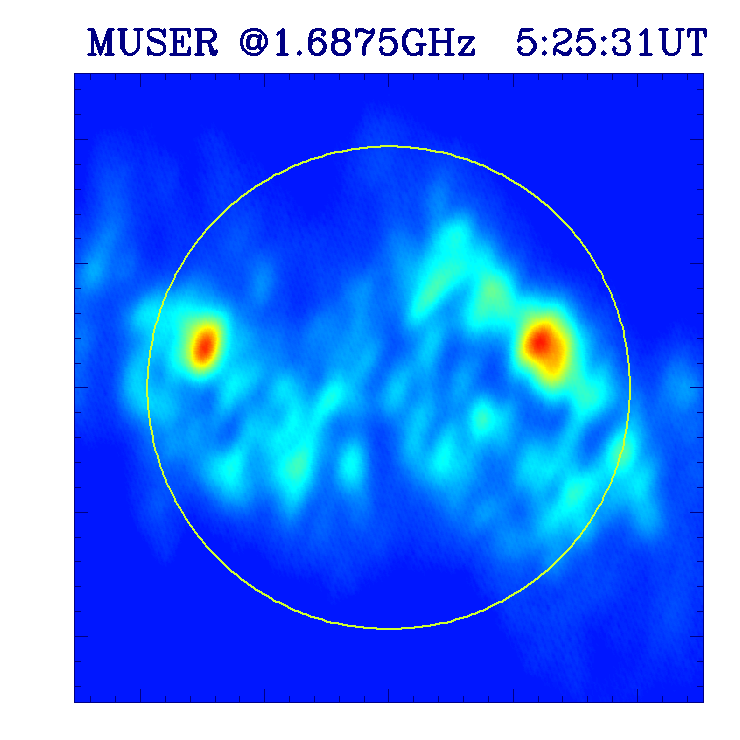}\\
\caption{The MUSER images with taking into account the rotation of solar $P$ angle at (a) 1.7125 GHz and (b) 1.6875 GHz on November 22, 2015 at 05: 25:31 UT after the position calibration by the present method.    }
  \label{MUSER1}
\end{figure}

\begin{figure}
  \centering
   \includegraphics[width=0.45\textwidth]{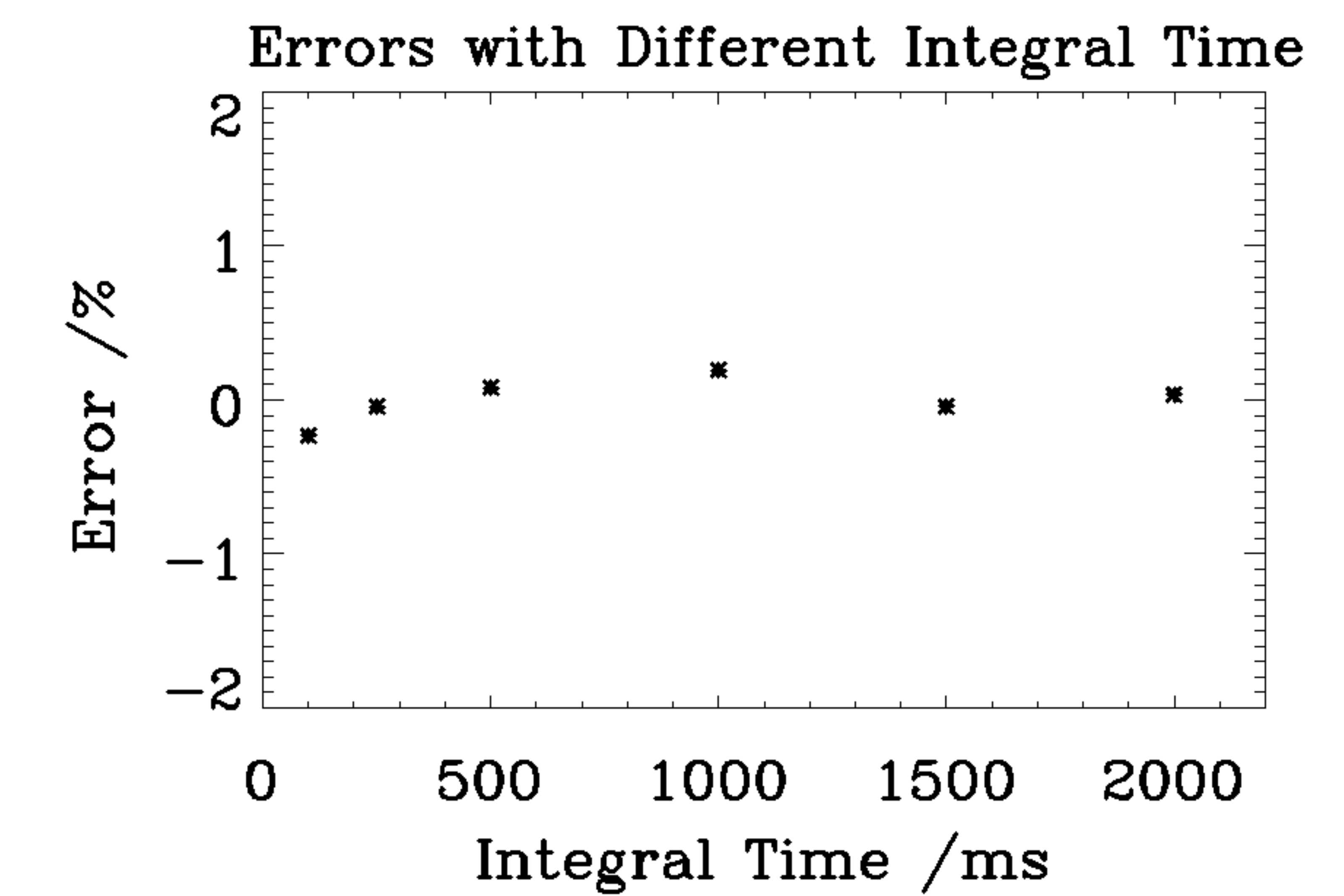}
  \includegraphics[width=0.45\textwidth]{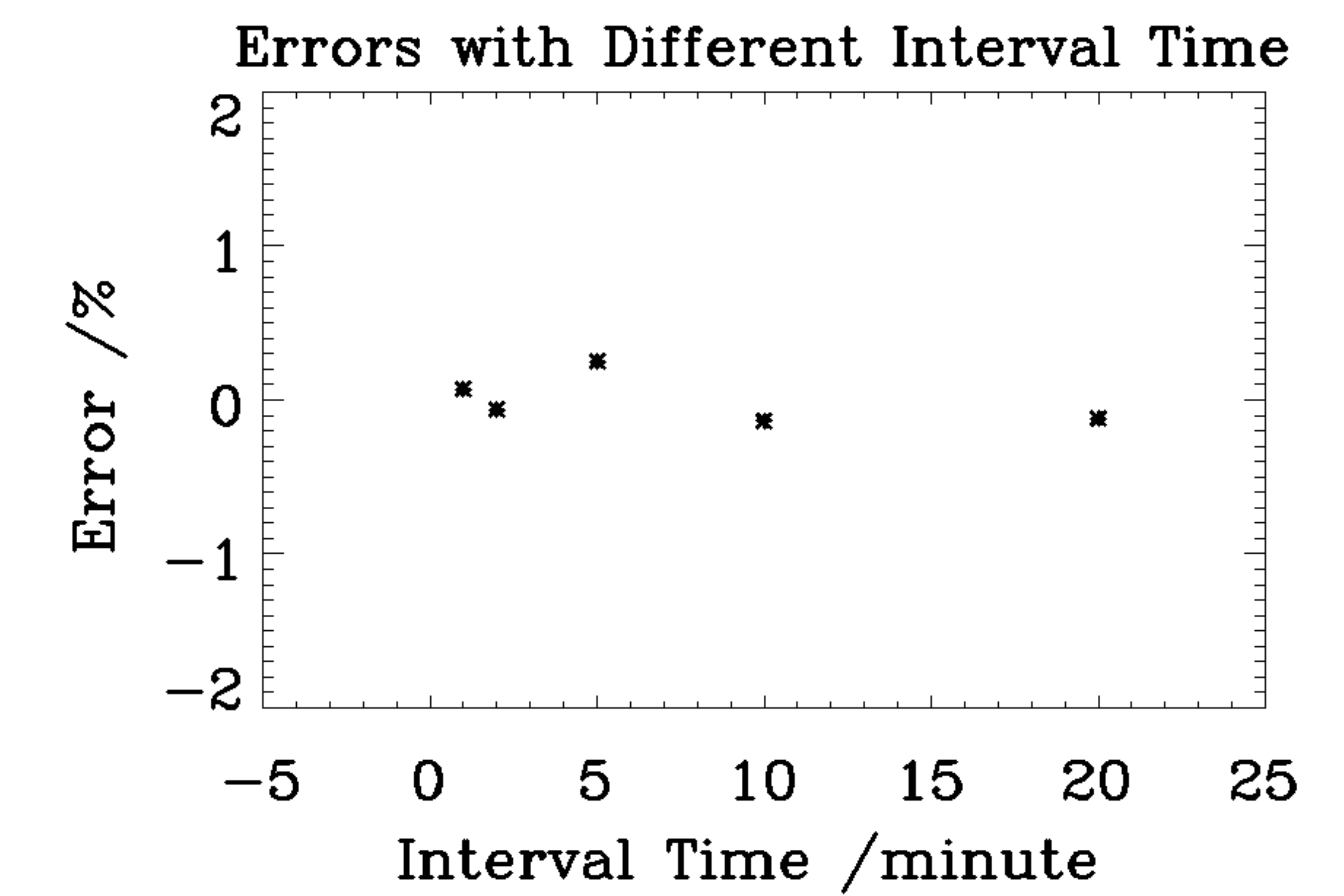}\\

  \caption{ Phase center relative errors changing with time interval and time cadence of images. (a) The results based on images with an integration time of 100 ms,  250 ms, 1000 ms, 1500 ms and 2000 ms, respectively. (b) The results based on images with a time cadence of 1min, 2 min, 5 min, 10 min and 20 min, respectively.}
  \label{sim2}
\end{figure}

To evaluate the effect of integration time and time cadence on final results, we calculate the variations of the restored phase center relative errors based on the  images with different integration time and different time cadence, respectively. The corresponding results are shown in Fig.~\ref{sim2}. The results in Fig.~\ref{sim2} (a) show that the relative errors fluctuate around -0.2\%$\sim$0.3\% when the integration time changes from 100 ms to 2000 ms. Similar results are achieved as shown in Fig.~\ref{sim2} (b) when the time interval varies from a few minutes to 20 minutes. In general, the relative errors are within 0.5 percent which indicate that both integral time and time cadence do not influence the restored result significantly.  This allows flexibility in selecting the appropriate observational intervals and the integration time whenever the reference source is stable.

\begin{figure}
  \centering
   \includegraphics[width=0.95\textwidth]{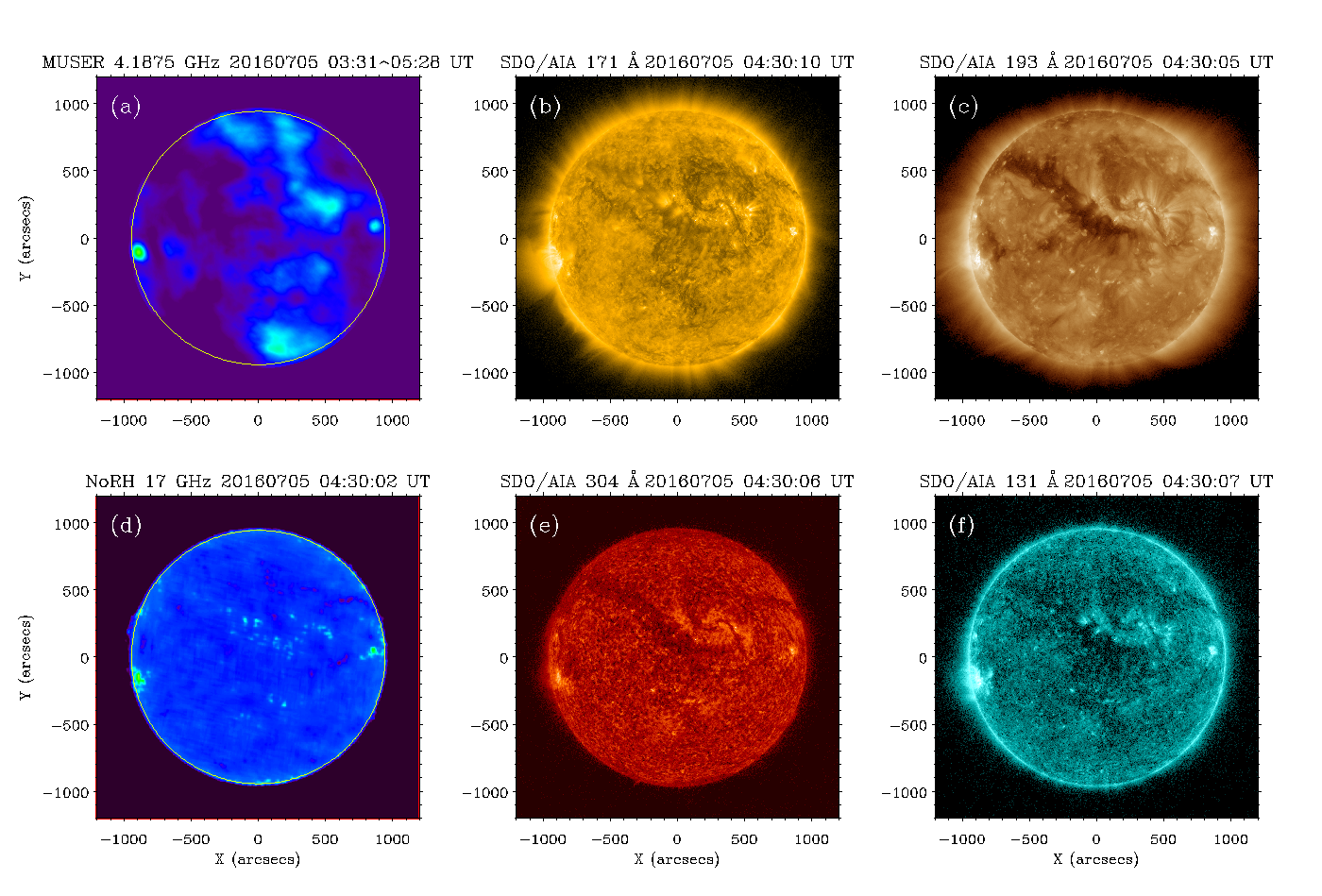}\\
\caption{(a) The MUSER-II image at 4.1875 GHz with taking into account the rotation of solar $P$ angle on July 5, 2016 synthesized from 03:31 UT to 05: 28 UT after the position calibration by the present method. (b) SDO/AIA 171~\AA~ EUV image at 04:30:10 UT.  (c) SDO/AIA 193~\AA~EUV image at 04:30:05 UT. (d)  NoRH 17 GHz image at 04:30:02 UT. (e) SDO/AIA 304~\AA~ EUV image at 04:30:10 UT. (f) SDO/AIA 131~\AA~ EUV image at 04:30:10 UT. }
  \label{MUSER2}
\end{figure}

MUSER-II observational data for the  quiet sun on July 5, 2016 were also calibrated, using another Geosynchronous satellite operating at around 4 GHz as a calibrator. There were no radio bursts that day. The  radio source observed by MUSER-II near east limb of the solar disk was selected as the reference source in this case.  The same iteration approach as in earlier applications was used, and the solar images were thus recovered, with the effects caused by deviation from the phase tracking centre eliminated by the current method. 
The MUSER-II synthesized image from 03:31 UT to 05:28 UT at 4.1875 GHz on July 5, 2016 after the position calibration is shown in Fig.~\ref{MUSER2}. For comparison the SDO/AIA EUV images at 171\AA (b), 193\AA (c), 304\AA (e) and 131\AA (f), and NoRH radio image at 17 GHz (d) around 04:30 UT are also shown in Fig.~\ref{MUSER2}. It can be seen that the observed features are in close agreement. 

Therefore, the results obtained by the proposed method in both simulations and using realistic observational data are satisfactory. They also demonstrate the desired performance of MUSER.

\section{Discussions and Conclusions}
\label{section5}

A general method has been proposed to calibrate the solar image position errors arising from calibrator offsets from the phase tracking center. 
For example, when the geosynchronous satellites are used in MUSER calibrations the present method can effectively resolve the problems of satellite deviation from the nominal phase reference position or phase tracking center. However, the currently available frequencies from satellites do not cover the full frequency bands used for MUSER observations. For the imaging at other frequencies, we need to seek some strong radio sources or intensive radio bursts on the sun as a calibrator source \citep{Wang+2013}. As to the amplitude calibration, the working frequency of geosynchronous satellites is usually narrow-band whereas  each bandwidth of MUSER frequency reaches 25~MHz, i.e., much wider than the artificial signal. Therefore, it is not possible to make use of observing artificial geosynchronous satellite for the amplitude calibration of MUSER visibility. Again, the standard techniques \citep{Bastian1989,Wang+2013,Mei+2017,Thompson+2017} including self-calibrations have been employed to calibrate the relative amplitude of MUSER images. Meanwhile, larger antennas for MUSER calibration system have been under construction. In the next a couple of years, two 20-meter 400 MHz - 2 GHz antennas and one 16-meter 2 GHz -15 GHz antenna with a He-cooled receiver will be incorporated into MUSER arrays for calibrations \citep{Yan+2021}.

\begin{figure}
  \centering
        \includegraphics[width=0.95\textwidth]{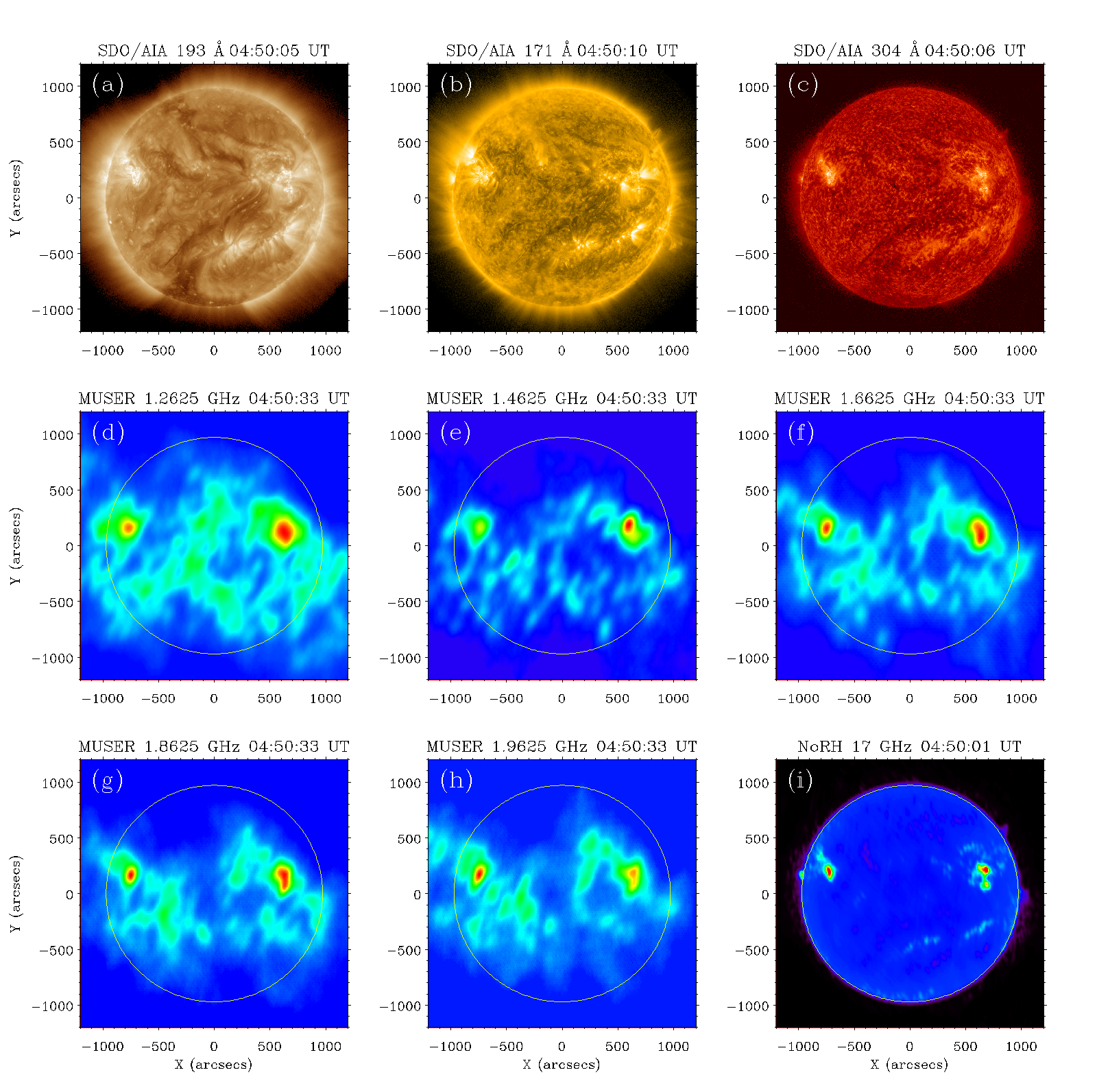}
    \caption{Some solar observations on November 22, 2015: (a) SDO/AIA 193 \AA ~EUV image at 04:50:05 UT, (b) SDO/AIA 171 \AA ~EUV image at 04:50:10 UT, (c) SDO/AIA 304 \AA ~EUV image at 04:50:06 UT, (d-h) the multi-frequency solar images observed at 04:50:33 UT by MUSER at 1.26 GHz, 1.46 GHz, 1.66 GHz, 1.86 GHz and 1.96 GHz, respectively, (i) the solar image observed at 04:50:01 UT by Nobeyama Radioheliograph at 17 GHz.}
    \label{MUSER3}
\end{figure}

As described in the previous section there was a C5.6 class flare on November 22, 2015 that peaked at 05:38 UT. For an impulsive radio burst on the sun, if we assume the radio burst originated from a compact area at its onset, we may treat the source as a $\delta$-function for phase calibration. During the observation of the sun, the phase center is always the center of the solar disk. Now we choose a source that deviates from the phase center as the calibrator. Then we can apply the present method to eliminate the deviation from the phase tracking center or the image center with the help of the observational data during the quiet period not severely affected by the flare. The model for the frequency-dependent solar radius in \citep{Tan+2015} is adopted in this method for calculating theoretical trajectories for frequencies from 0.4 GHz to 15 GHz. Fig.~\ref{MUSER3} shows the restored multi-frequency images from MUSER-I at  1.26 GHz, 1.46 GHz, 1.66 GHz, 1.86 GHz and 1.96 GHz and the comparison with observations in other wavelengths such as EUV images from SDO/AIA and the radio image from the Nobeyama Radioheliograph at 17 GHz. These results indicate that MUSER observations are both reliable and significant in revealing solar atmospheric observational features. The goal of this research is to demonstrate that the proposed new position calibration model and solution technique are reliable. The interpretation of MUSER radio images that have been restored will be presented elsewhere (e.g., \citealt{Chen+2019,Zhang+2021}).

In summary, the conclusions are as follows. 

1. A mathematical formula describing the phase calibration problem in the aperture synthesis images arising from a calibrator position offset from the phase tracking center has been established.  According to the aperture synthesis principle,  the phase tracking center is the calibrator position and it is at the origin of the sky radio image plane. If a calibrator offset (with either a known or unknown value) from the phase tracking center is employed in the phase calibration, this offset will be transferred to the sky radio images. Then it is shown, for the first time, that the observed dirty image of the sky radio intensity distribution can be formulated explicitly  as a convolution product between a shifted sky radio image with unknown deviation and a {blurring function, as  expressed in equation~({\ref{eq:sunshift}}). This blurring function has a modulus of unity and approaches a $\delta$-function when the deviation reduces to zero. Therefore, the shifted sky radio image is merely modulated by the introduced blurring function, and it becomes the correct sky radio image as the deviation goes to zero. The newly derived mathematical formula can also be applied to other synthesis imaging analyses.}

2. The corresponding position calibration procedure has been proposed to determine the calibrator offset from the phase tracking center based on the above mentioned formula by investigating the offset of the positions of radio images over a certain period of time. This is achieved by selecting a stable radio source in the field of view as a reference spot. Then, the reference source position with respect to its original non-deviated phase tracking center should follow its original geometric relationship with respect to the non-deviated origin during the observation interval, e.g., a stable spot on the sun will vary its position solely due to the solar rotation, or a radio source in the sky map will just keep its position unchanged. This constitutes the criterion for the proposed optimization model of the new position calibration procedure, e.g., equation~(\ref{eq:model}) for the solar observations. {Simulation tests show that the proposed method can effectively eliminate errors due to known or unknown calibrator offsets from the phase tracking centre to small fractions of the corresponding angular resolution under a variety of conditions with different noise levels and sampling configurations}. This demonstrates that the proposed new position calibration method is valid and practical.

 3. MUSER observational data have been treated by the proposed method and the calibrated results are robust under different  integration time and cadences. When the restored MUSER radio images are compared to other solar observations, it can be seen that the mutual co-alignments agree in exhibiting the observed features in these images, supporting the calibration and MUSER's desired performance. Scientific discussion of the MUSER observations, however, will be given elsewhere. The present study contributes to MUSER calibration, and the future update of the MUSER calibration system will enable MUSER to be used in a broader range of solar and heliospheric physics applications.

\section*{Acknowledgements}

This work was supported by NSFC grants (11790301, 11790305, 11773043, U2031134, 12003049), and National Key R\&D Program of China  (2021YFA1600500, 2021YFA1600503, 2018YFA0404602). MUSER was supported by National Major Scientific Research Facility Program of China with the grant Number ZDYZ2009-3. The MUSER calibration system is a part of the Chinese Meridian Project funded by China's National Development and Reform Commission. The HMI/SDO magnetogram, AIA/SDO and NoRH images are obtained from their respective web sites which are sincerely acknowledged. NoRH was operated by ICCON from 2015 to 2020. We thank the MUSER Team for MUSER operation. Mr. Tulsi Thapa is acknowledged for modifying the English. Dr. Tim Bastian is greatly appreciated for reading and improving the English of the manuscript.

\bibliographystyle{spmpsci}     


\label{lastpage}

\bibliographystyle{raa}
\bibliography{arViv}

\end{document}